\documentclass[floatfix,nofootinbib,twocolumn,preprintnumbers,superscriptaddress,notitlepage,nolongbibliography]{revtex4-2}

\usepackage{times}
\usepackage{xcolor}
\colorlet{RED}{red}
\usepackage{soul}
\usepackage{listings}
\usepackage{booktabs}
\usepackage{amsfonts}
\usepackage{textcase}
\usepackage{amsmath}
\usepackage{multirow}
\usepackage{amssymb}
\usepackage{bm}
\usepackage{dcolumn}
\usepackage{enumerate}
\usepackage{epsfig}
\usepackage{graphicx}

\usepackage{graphics}
\usepackage[latin1]{inputenc}
\usepackage{latexsym}
\usepackage{placeins}  
\usepackage{capt-of}   

\usepackage{rotating}
\usepackage{hyperref}
\usepackage[normalem]{ulem}
\usepackage{orcidlink}

\usepackage[T1]{fontenc}
\usepackage{upquote}
\usepackage{etoolbox}

\robustify{\texttt}
\let\originaltexttt\texttt

\begingroup
\catcode`'=\active
\catcode``=\active
\makeatletter
\renewrobustcmd{\texttt}[1]{%
   {%
   \everyeof{\noexpand}\endlinechar-1
   \expandafter\catcode\string``=\active
   \expandafter\catcode\string`'=\active
   \let'\textquotesingle
   \let`\textasciigrave
   \ifx\encodingdefault\upquote@OTone
    \ifx\ttdefault\upquote@cmtt
     \def'{\char13 }\def`{\char18 }%
    \fi
   \fi
   \scantokens{\originaltexttt{#1}}%
   }%
}%
\endgroup

\definecolor{bg}{rgb}{0.95,0.95,0.95}
\definecolor{keyword}{rgb}{0,0,0.8}
\definecolor{string}{rgb}{0.8,0,0}
\definecolor{comment}{rgb}{0,0.5,0}

\newcommand{\<}{\begin{equation}}
\newcommand{\?}{\end{equation}}

\definecolor{mgreen}{rgb}{0.1,0.7,0.1}

\newcommand{\edit}[1]{{\textcolor{black} {#1}}}

\begin{document}

\title{Tracing the evolution of eccentric precessing binary black holes: a hybrid approach}
\author{Amitesh Singh \orcidlink{0000-0002-4697-1254}}
\email{amiteshsingh487@gmail.com}
\affiliation{Department of Physics and Astronomy, The University of Mississippi, University, Mississippi 38677, USA}

\author{Nathan~K.~Johnson-McDaniel \orcidlink{0000-0001-5357-9480}}
\email{nkjm.physics@gmail.com}
\affiliation{Department of Physics and Astronomy, The University of Mississippi, University, Mississippi 38677, USA}

\author{Anuradha Gupta \orcidlink{0000-0002-5441-9013}}
\email{agupta1@olemiss.edu}
\affiliation{Department of Physics and Astronomy, The University of Mississippi, University, Mississippi 38677, USA}

\author{Khun Sang Phukon \orcidlink{0000-0003-1561-0760}}
\email{k.s.phukon@bham.ac.uk}
\affiliation{School of Physics and Astronomy and Institute for Gravitational Wave Astronomy,\\University of Birmingham, Edgbaston, Birmingham, B15 2TT, United Kingdom}

\date{\today}

\begin{abstract}

To describe a general bound binary black hole system, we need to consider orbital eccentricity and the misalignment of black holes' spin vectors with respect to the orbital angular momentum. While binary black holes produced through many formation channels have negligible eccentricity close to merger, they often have a non-negligible eccentricity at formation, and dynamical interactions could produce binaries with non-negligible eccentricity in the bands of current and proposed gravitational-wave (GW) detectors. Another quantity that carries information about the formation channel is the angle between each black hole's spin vector and the binary's orbital angular momentum (referred to as the spin tilt) at formation. The spin tilts inferred in GW astronomy are usually those when the binary is in the band of a GW detector, but these can differ significantly from those at formation. Therefore, it is necessary to evolve the binary back in time to compute the tilts at formation. For \edit{certain binaries in} formation scenarios \edit{where the orbital angular momentum at formation is large,} the tilts in the formal limit of infinite orbital angular momentum, also known as tilts at infinity, \edit{could be a useful} approximation to those at formation. \edit{Additionally, full orbit-averaged evolution to formation could be computationally infeasible in many situations, while the hybrid evolution we use to compute the tilts at infinity would be feasible. Here, the tilts at infinity provide a natural common reference for comparing observational constraints on spin distributions with the predictions from population-synthesis models, since the tilts at infinity include all the information about spin orientations accessible via precession-averaged evolution.} We thus generalize the publicly available \texttt{tilts\_at\_infinity} code to compute the tilts at infinity for eccentric, spin-precessing binaries. This code employs hybrid post-Newtonian evolution, starting with orbit-averaged evolution for higher frequencies and then transitioning to precession-averaged evolution to compute the tilts at infinity. We find that the transition frequency used in the quasicircular case still gives acceptably small errors in the eccentric case, and show that eccentricity and hybrid evolution both have a significant effect on the tilts at infinity for many binaries. In particular, we give examples of cases where the tilts at infinity are and are not a good approximation to the tilts at formation in the eccentric case.

\end{abstract}

\maketitle

\section{Introduction}
\label{sec:intro}

Binary black hole (BBH) detection is now a fairly frequent occurrence, thanks to the recent improvements in detector sensitivity and other advancements in hardware and software for the current ground-based LIGO and Virgo detectors~\cite{LIGOScientific:2014pky,VIRGO:2014yos}.
The LIGO-Virgo-KAGRA Collaboration (LVK) has detected more than \edit{$300$} compact binary coalescences so far~\edit{\cite{LIGOScientific:2026wfs}}, including BBHs, binary neutron stars, and black hole-neutron star binaries. This number is expected to increase rapidly as the current detectors (including KAGRA~\cite{KAGRA2021}) continue to increase their sensitivity~\cite{Aasi:2013wya} and next generation detectors come online, both ground-based detectors like Einstein Telescope~\cite{punturo2010einstein} and Cosmic Explorer~\cite{Evans:2021gyd} as well as space-based detectors like LISA~\cite{LISA:2024hlh}.

Understanding the mechanisms that form the observed BBHs (see, e.g.,~\cite{Mapelliformation2021}) is a major goal of gravitational-wave (GW) astronomy, and the distributions of the binaries' inferred parameters (e.g., masses, spin vectors, eccentricities, and luminosity distances) provide important clues about these mechanisms (e.g.,~\cite{Zevin:2017evb,Zevin:2021rtf,Bouffanais:2021wcr,PhysRevD.106.103013,Colloms:2025hib,Afroz:2025ikg,Berti:2025usa}).
For instance, one can infer which black hole formed first in a binary~\cite{Mould_2022}, and the properties of the progenitor galaxies of the BBHs~\cite{Srinivasan:2023vaa}. See~\edit{\cite{LIGOScientific:2026ctl}} for the latest LVK population inference results, using both parametric and nonparametric methods.
In particular, the distribution of spin tilts (polar  angles of spins measured from the orbital angular momentum) is crucial to distinguishing between different BBH formation channels. An isotropic distribution of black hole spins is indicative of dynamical formation, while aligned spins suggest an isolated formation process (see~\cite{Stevenson:2017dlk}). Misaligned spins lead to a rich phenomenology of spin precession---see, e.g.,~\cite{Apostolatos:1994mx,Schnittman:2004vq,Gerosa:2015tea,Gerosa:2018mwg,Gangardt:2021lic}---and specific precessional signatures can be related to specific formation scenarios, e.g.,~\cite{Gerosa_2018,steinle2022signatures}. There are analyses looking for such signatures in individual BBHs observed using GWs~\cite{gangardt2022constraining,Johnson-McDaniel:2023oea} and in the population~\cite{Varma:2021xbh,Biscoveanu:2025jpc}, though there is not yet significant evidence for any of them, even though there are several binaries with evidence for precession---see~[\edit{\citealp{LIGOScientific:2026wfs}}, \citealp{LIGOScientific:2025brd}] for the most recent results---as well as evidence for misaligned spins in the population~\edit{\cite{LIGOScientific:2026ctl}}.

In addition to precessional dynamics, orbital eccentricity also plays an important role in distinguishing binary formation channels. Over long timescales, GWs efficiently reduce orbital eccentricity by extracting angular momentum from the system~\cite{PhysRev.136.B1224}. Consequently, binaries formed through isolated channels are typically expected to circularize before entering the detection band of current ground-based GW detectors. Conversely, some binaries originating from dynamical channels, such as active galactic nuclei or globular clusters, are anticipated to retain non-negligible eccentricities at detection frequencies~\cite{Romero-Shaw:2021ual}. Over the last few years, there have been claims that some of the observed BBHs could be eccentric~\cite{Gayathri:2020coq,Romero-Shaw:2020thy,Romero-Shaw:2021ual,Romero-Shaw:2022xko,Gupte:2024jfe,Planas:2025jny}. With the advent of upcoming detectors like LISA and next-generation terrestrial GW detectors, the prospects of detecting eccentric BBHs will only keep increasing. \edit{Reference}~\cite{Fumagalli:2024gko} shows that if one neglects a small but nonzero residual eccentricity at $10$ Hz ($e_{\rm 10\, Hz} \le 0.05$), it can lead to significant biases in the spin tilts when evolving the binary backwards to a large separation ($\sim 10^{4}\edit{M}$, \edit{where $M$ is} the binary's total mass in geometrized units) to find the spin tilts at a point closer to the binary's formation. Hence, we need to be prepared for a possible observation of a precessing binary with non-negligible eccentricity and need to be able to compute the tilts at infinity (i.e., the tilts in the formal limit of infinite angular momentum, discussed further below) in such cases.

The measurements of spin tilts in the third observing run were not very precise, with no significant difference between the posterior distributions of spin tilts at infinity and at the reference frequency for either the individual events~\cite{Johnson-McDaniel:2021rvv} or for the population~\cite{Mould:2021xst}. However, GW241011~\cite{LIGOScientific:2025brd}, which was detected during the fourth observing run, has a better constrained primary spin tilt measurement, though the posteriors at the reference frequency and those at infinity just differ by a small shift and still mostly overlap. Regardless, the precision of spin tilts measurements will improve with the sensitivity of the detectors, e.g., in the plus-era of detectors~\cite{Knee:2021noc, Kulkarni:2023nes}. In particular,~\cite{Kulkarni:2023nes} showed that it is possible to measure the tilts at detection frequency with an accuracy of $0.06~\text{rad}$ (with $90\%$ credible interval) for certain comparable-mass quasicircular binaries with the plus-era sensitivity of the LIGO and Virgo detectors (expected to be attained during the fifth observing run~\cite{Aasi:2013wya}). Thus, for the third-generation detectors like Cosmic Explorer and the Einstein Telescope, which are $\sim 10$ times more sensitive than the plus-era detectors, it may be possible to measure these tilts with an accuracy of $\sim 0.006~\text{rad}$ for certain binaries. (\edit{Reference}~\cite{Klein:2022rbf} finds an accuracy of $0.02~\text{rad}$ for Cosmic Explorer alone for one case, though with different parameters, notably a somewhat smaller total mass and spins, than the binary considered in~\cite{Kulkarni:2023nes} that gives the best accuracy.)

\edit{Reference}~\cite{Pratten:2023krc} predicts that LISA will be able to measure the primary tilt with a $90\%$ credible interval of width $\sim0.003$~rad in some cases with a mass ratio of $q = 4$. In contrast, the tightest $90\%$ credible intervals predicted for binaries with a mass ratio of $q \simeq 1.4$ and somewhat larger spin magnitudes \edit{in}~\cite{2025PhRvD.112f4017G} have a width of $\sim0.012$~rad. \edit{There are also estimates \edit{in}~\cite{klein2016science}, but they have a few caveats, including the use of inspiral-only waveforms and the old LISA detector model}.

The tilts inferred in the detectors' sensitive band can be significantly different from those at infinity, as is illustrated in, e.g.,~\cite{Renzis_2023,Kulkarni:2023nes}. \edit{(References~\cite{Johnson-McDaniel:2021rvv, Gerosa:2015tea} illustrate these differences without including the statistical measurement uncertainties from finite detector sensitivities, \edit{while}~\cite{Apostolatos:1994mx} was the first to study the evolution of the spin tilts due to the BBH inspiral.)} Additionally, considering specific \edit{population-synthesis} predictions without the \edit{statistical measurement} uncertainties from finite detector sensitivities, one finds that there can be significant differences ($> 0.1$, and even $> 1$ in some cases) between the cosines of the tilts at formation and $10$~Hz \edit{in various formation scenarios}. Specifically, as illustrated in Appendix~\ref{app:pop_syn_evol}, we consider both dynamical and \edit{isolated} formation channels:  For dynamical formation, we consider (i) three-body induced binaries (binaries assembled and hardened through binary-single three-body resonant encounters in dense stellar systems); (ii) remnant binaries from the von Zeipel-Lidov-Kozai (ZLK) process~\cite{von1910application,Kozai:1962zz,Lidov:1962wjn} in hierarchical triples (where a distant tertiary induces secular eccentricity-inclination oscillations that drive the inner binary to high eccentricity and a GW driven merger); and (iii) binary-single captures, using the predictions from~\cite{Kritos:2022ggc}. For isolated formation, we use the predictions from~\cite{Belczynski:2017gds}. \edit{All the dynamical and isolated BBHs we study in this work merge within a Hubble time.}

edit{Reference}~\cite{Johnson-McDaniel:2021rvv} showed that spin tilts at infinity are a good approximation to those at formation for many formation scenarios.
The spin tilts at infinity are obtained in the limit where the binary's orbital angular momentum $L \to \infty$, since in this limit the amplitude of spin precession goes to zero. This limit is attained theoretically for exactly quasicircular binaries, though in practice there is always a finite uncertainty in the tilts at formation from two-spin precession (proportional to $1/L$), unless one knows exactly when in the binary's precessional cycle it was formed, even though in principle (albeit not in practice) this uncertainty can be arbitrarily small. However, for eccentric binaries, the limit of infinite orbital separation corresponds to the limit where the eccentricity $e \nearrow 1$, and the orbital angular momentum approaches a finite value, as can be seen from the expressions in~\cite{PhysRev.136.B1224}, as discussed \edit{in}~\cite{Johnson-McDaniel:2021rvv,Fumagalli:2023hde}. Nevertheless, we still compute the tilts at infinity for eccentric binaries: The precession-averaged evolution~\cite{Johnson-McDaniel:2021rvv} we use to perform this computation is insensitive to the presence of eccentricity, and the tilts at infinity still provide a well-defined (idealized) reference point for these cases. Moreover, in some eccentric formation scenarios, $L$ at formation will still be large enough that the \edit{differences between the tilts at infinity and those at formation} will be small, as discussed in Sec.~\ref{sec:inf_vs_form}.
However, dynamically formed binaries with extremely high eccentricities at formation ($e \simeq 0.99999999$, or $\edit{1 - e \simeq}\, 10^{-8}$), which lead to a significant eccentricity in the band of ground-based detectors, can possess much lower angular momentum values, as low as $L/M^2 \simeq 0.1$ in geometrized units (using the \edit{population-synthesis} data from~\cite{Kritos:2022ggc}). In those cases, one will not want to compute the tilts at infinity as an approximation to the tilts at formation, but may rather be able to compute the tilts at formation to a good approximation by evolving backwards to a finite (and not particularly large) separation, as we will show with \edit{some examples in Sec.~\ref{sec:inf_vs_form}}.

\edit{The tilts at infinity will also be useful in cases where full orbit-averaged evolution between formation and the detector band (or vice versa) is computationally infeasible and only hybrid evolution is practical. Because the precession-averaged portion of a hybrid evolution does not retain the instantaneous precessional phase, evolving detector-band posterior samples backward from the reference frequency to a finite formation point determines the tilts only up to their variation over a precessional cycle. Conversely, evolving binaries from population-synthesis models forward requires sampling the precessional phase when transitioning to orbit-averaged evolution. Tilts at infinity provide a natural common reference for the precessional information retained by the precession-averaged evolution, so population-synthesis predictions and detector-band posterior samples can be compared in the same reference variables. We discuss this argument in detail in Sec.~\ref{sec:concl} and illustrate the resulting distributions of tilts at infinity for selected formation scenarios in Appendix~\ref{app:tilts_at_infty_specific_formations}.}

There exist codes to evolve quasicircular, precessing BBHs to infinite separation~\cite{Johnson-McDaniel:2021rvv, gerosa2023efficient}.
However, there are not yet any public codes for evolving eccentric, precessing binaries to infinity. It is important to have such a code, particularly to ensure that the systematic errors in the computation of spin tilts at infinity remain smaller than the anticipated statistical \edit{measurement} uncertainties for future GW observations.
Thus, we extend the \texttt{tilts\_at\_infinity} code~\cite{Johnson-McDaniel:2021rvv} in the LALSuite GW data analysis package~\cite{LALSuite} to evolve eccentric, precessing BBHs in a hybrid approach: we use an orbit-averaged post-Newtonian (PN) evolution code~\cite{Phukon:2025yva} (hereafter referred to as the orbit-averaged code) at small orbital separations and switch to faster but less accurate precession-averaged equations~\cite{Johnson-McDaniel:2021rvv} at large orbital separations, where the precession-averaged evolution is sufficiently accurate. This hybrid approach provides a more accurate way to compute the tilts at infinity than just using precession-averaged evolution as in~\cite{Fumagalli:2023hde}. While we wait for this code to be reviewed by the LVK collaboration to be officially included in LALSuite master~\cite{LALSuite}, an unreviewed version of the code is publicly available at~\cite{amiteshgit} via a fork of LALSuite.

We find that using the same transition orbital velocity (depending only on the binary's mass ratio) as that obtained in the quasicircular case in~\cite{Johnson-McDaniel:2021rvv} gives acceptably small errors in the tilts at infinity due to the choice of the transition orbital velocity. Here the errors are computed using the difference in the spin tilts at infinity obtained using the empirically determined transition orbital velocity as in~\cite{Johnson-McDaniel:2021rvv}, and half of that value. These errors are all lower than $10^{-2}$ in the cosines of the tilts at infinity and thus smaller than the anticipated statistical errors for observations in the upcoming fifth observing run of the advanced GW detector network~\cite{Kulkarni:2023nes}. We do not demand the same accuracy of $10^{-3}$ that is achieved for the quasicircular code~\cite{Johnson-McDaniel:2021rvv}, since the eccentric orbit-averaged evolution is less accurate than the quasicircular one. In particular, the eccentric orbit-averaged evolution only includes the leading spin-orbit and spin-spin terms (i.e., $2$PN accuracy), while~\cite{Johnson-McDaniel:2021rvv} finds that in the quasicircular case the first PN correction to the spin-orbit terms (at $2.5$PN) leads to differences as large as $\sim 0.1$.\footnote{We use the standard PN order counting where $n$PN corresponds to terms of order $(v/c)^{2n}$ beyond leading order ($v$ is the binary's orbital velocity), except for the precession equations, where the order counting is such that the leading spin-orbit term appears at $1.5$PN as it does in the binary's orbital dynamics.} Thus, it will almost surely be necessary at least to extend the eccentric orbit-averaged code~\cite{Phukon:2025yva} to $2.5$PN (and possibly $3$PN) accuracy in the spins and $3.5$PN in the nonspinning terms (as we show later) in order to obtain results that have errors $< 10^{-2}$. 

As discussed in~\cite{Phukon:2025yva}, the extension of the orbit-averaged evolution to $3$PN spinning terms just requires the extension of the quasi-Keplerian formalism~\cite{AIHPA_1985__43_1_107_0} for arbitrary spin orientations to $2.5$PN, since the other ingredients are present in the literature. To extend this evolution to $3.5$PN in the non-spinning terms, one needs to compute the $3.5$PN contributions to the orbit-averaged fluxes for eccentric binaries, which has not yet been done.
However, the hybrid evolution code presented in this work already provides an improvement on the accuracy of the only precession-averaged eccentric evolution used in~\cite{Fumagalli:2023hde,Fumagalli:2024gko}, since we find that the differences in cosine tilts estimated from the only precession-averaged and hybrid evolutions can approach unity for some binaries.
In addition to studying the accuracy of the code, we also use it to illustrate how eccentricity affects the tilts at infinity and perform some initial investigations into how well the tilts at infinity approximate those at $e \nearrow 1$ in cases where eccentricity is not negligible.

The rest of the paper is organized as follows: In Sec.~\ref{sec:orbital_spin_evol}, we introduce the orbital and spin parameters for an eccentric, precessing BBH, and provide details about the evolution equations we use. In Sec.~\ref{sec:interface}, we discuss how we interface orbit-averaged and precession-averaged evolution equations to create a unified code
to compute the tilts at infinity for  precessing BBHs on eccentric orbits, and present various checks of the accuracy of the results. In Sec.~\ref{sec:spin_tilts_inf}, we discuss the effects of eccentricity on the tilts at infinity, and compare the hybrid evolution with purely precession-averaged evolution. Then in Sec.~\ref{sec:inf_vs_form} we give examples of where the spin tilts at infinity are good and poor approximations to those at $e \nearrow 1$ as well as at formation for some dynamically formed binaries. Finally, we summarize and conclude in Sec.~\ref{sec:concl}. In Appendix~\ref{app:pop_syn_evol}, we compute the differences between the tilts at infinity at $10$~Hz for some \edit{population-synthesis models, while in Appendix~\ref{app:tilts_at_infty_specific_formations}, we illustrate how the distributions of tilts at infinity differ for some example isolated and dynamical formation scenarios.} Additionally, we give an example usage of the code in Appendix~\ref{app:example_usage} and derive a bound on the change in tilts between eccentricities close to $1$ and the limiting value of $1$ in Appendix~\ref{app:delta_theta_bound}. We use geometrized ($G = c = 1$) units, and express all angles in radians throughout this paper.
\edit{At various places in this work, we will refer to three kinds of uncertainties in the tilts: \textit{statistical uncertainties} in the measurement from finite detector sensitivities, the \textit{intrinsic uncertainties} due to the unknown precessional phase of the BBHs at formation, and the \textit{errors} from the choice of the transition frequency. These different uncertainties can differ significantly from each other, sometimes by orders of magnitude.}

\section{Orbital and spin evolution}
\label{sec:orbital_spin_evol}

\subsection{Orbit-averaged evolution}

Here, we discuss the parameters used to describe the orbital motion and spin-precession of eccentric BBHs, as reviewed in Sec.~II of~\cite{Phukon:2025yva}. We first discuss the orbital dynamics.
Let us consider a BBH with black hole masses \(m_1\) and \(m_2\), mass ratio \(q = m_2/m_1 \leq 1\), total mass $M = m_1 + m_2$, and dimensionless spin magnitudes of $\chi_1$ and $\chi_2$. The spin orientations are determined by the tilt angles $\theta^{}_{1,2}\in [0,\pi]$ between each spin and the binary's orbital angular momentum as well as the azimuthal angles $\psi^{}_{1,2} \in [0, 2\pi)$ between the projections of the spin vectors onto the orbital plane and the binary's line of periastron (see Fig.~1 in~\cite{Phukon:2025yva} for a visual representation of these angles and orbital elements). We use the quasi-Keplerian parameterization of the binary's orbital motion, which we express in terms of an average orbital frequency $\omega$ and the temporal eccentricity $e_t$. (We will refer to this eccentricity as just $e$ for the remainder of the paper, except where we want to emphasize its specific definition.) Later in this paper, we will also consider quasicircular binaries (with $e_{t}=0$) whose orbit-averaged evolution dynamics~\cite{SpinTaylor_TechNote} do not require the quasi-Keplerian formalism.

We also use the orbit-averaged dissipative dynamics (from radiation reaction) and spin-precession equations. Specifically, for the nonspinning terms, we use the $3$PN computations from~\cite{Memmesheimer:2004cv,Arun:2009mc} (the highest order for which the dissipative dynamics have been computed), along with the high-accuracy hyperasymptotic expansions for the eccentricity enhancement functions in the energy and angular momentum fluxes from Loutrel and Yunes~\cite{Loutrel:2016cdw}. The spinning contributions to the orbital dynamics are the $2$PN expressions from~\cite{Klein:2010ti,Klein:2018ybm}, the highest order to which this has been computed for arbitrary spin orientations (so we are including the leading-order spin-orbit and spin-spin interactions). The spin-precession equations (from~\cite{Racine:2008qv}) are also $2$PN accurate. In all of these expressions, the spin-induced quadrupole moments are specialized to their black hole values. The code~\cite{Phukon:2025yva} to evolve orbit-averaged evolution makes various internal consistency checks that the PN equations are not obviously incorrect (listed in Table~I of~\cite{Phukon:2025yva}) and some of these can be triggered when evolving backwards in the hybrid evolution to compute the tilts at infinity, particularly due to the large eccentricities one obtains, as we will discuss later.

\subsection{Precession-averaged evolution}

To efficiently evolve BBHs over long timescales,~\cite{Kesden:2014sla, Gerosa:2015tea} introduced the precession-averaged approach. In particular, this allows one to compute the tilt angles at infinity (specifically infinite orbital angular momentum), which are well-defined except in the exactly equal-mass case, where the tilts do not approach a single value in that limit (see, e.g., the discussion in Sec.~III~A of~\cite{Johnson-McDaniel:2021rvv} as well as~\cite{Gerosa:2016aus}). The precession-averaged approximation relies on the precession timescale being much smaller than the radiation reaction timescale and is only $1$PN accurate in the dissipative dynamics, as discussed in~\cite{Gerosa:2015tea}. Thus, the precession-averaged dynamics lose accuracy close to the BBH merger, where we instead use the orbit-averaged evolution.

As discussed in Sec.~II~A of~\cite{Johnson-McDaniel:2021rvv}, following~\cite{Yu:2020iqj}, if one just wants to compute the tilts at infinity for eccentric binaries, there is no change to the precession-averaged equations compared to the quasicircular case; see also~\cite{Fumagalli:2023hde}. (Of course, the eccentricity increases rapidly when evolving backwards, so the orbit averaging that underlies the precession-averaged evolution will become less accurate, due to the significant difference in orbital speeds at periastron and apastron. However, this caveat also applies to the orbit-averaged equations we use.) Thus, we use the same precession-averaged code as in~\cite{Johnson-McDaniel:2021rvv}. The only difference compared to the quasicircular case is that we use the eccentric expression for the magnitude of the orbital angular momentum to initialize the precession-averaged evolution. Here we use the Newtonian order expression, since this is what enters into the $2$PN precession equations used in the orbit-averaged evolution:
\begin{equation}\label{eq:L}
	L = m_1m_2\frac{\sqrt{1 - e^2}}{v},
\end{equation}
where $v = (M\omega)^{1/3}$ is the orbital velocity corresponding to the orbit-averaged orbital angular frequency $\omega$ (as used in the orbit-averaged code~\cite{Phukon:2025yva}).

\section{Interfacing orbit-averaged and precession-averaged evolutions}
\label{sec:interface}

A key element of the hybrid method we use is determining the transition orbital velocity $v_{\rm trans}$ where we change from using orbit-averaged to precession-averaged evolution equations. Since the orbital velocity varies over an orbit in the eccentric case, even with just the conservative dynamics, we use the orbital velocity obtained from the orbit-averaged angular velocity to set the transition orbital velocity.
The specific value of $v_{\rm trans}$ one uses depends on the accuracy one requires for the resulting tilts at infinity. The quasicircular hybrid evolution code in~\cite{Johnson-McDaniel:2021rvv} uses a $v_{\rm trans}$ value obtained by demanding an accuracy of better than $10^{-3}$ in the cosines of the tilts at infinity, motivated by the expected measurement accuracy with next-generation GW detectors. While one might expect that the appropriate $v_{\rm trans}$ in the eccentric case would depend on the eccentricity, we find that using the quasicircular $v_{\rm trans}$ gives acceptable accuracy in the eccentric case.

The transition velocity in the quasicircular case is given by
\begin{equation}
	v_{\rm trans} = -0.05q^2 + 0.06.
	\label{eq:vtrans_vs_q}
\end{equation}
In the quasicircular case, the ($m = 2$ mode) GW frequency is given by $f_{\rm trans} = v_{\rm trans}^{3}/(\pi M)$. We use the same expression in the eccentric case, since the orbit-averaged code sets $f_\text{GW} = \omega/\pi$, where $\omega$ is the orbit-averaged orbital frequency. Thus, we are determining the transition point in terms of the orbit-averaged orbital frequency, as mentioned above. To validate their $v_{\rm trans}$ on binary systems different from the set used to obtain the expression,~\cite{Johnson-McDaniel:2021rvv} generated a synthetic set of $500$ binary systems, with parameters drawn from the following ranges: $m_1, m_2 \in [5, 100]M_\odot$; $\chi_1, \chi_2 \in [0,1]$; $\theta_1, \theta_2 \in [0, \pi]$; $\phi_{12} \in [0, 2\pi]$,   where $\phi_{12}$ is the  difference between the projections of the spin vectors onto the orbital plane. The spin angles are specified at a  reference frequency $f_{\text{ref}} = 20$~Hz.
For each binary, they computed the absolute difference in the cosines of the tilts at infinity $\cos\theta_A^\infty$ ($A\in\{1,2\}$) obtained using the $v_{\text{trans}}$ [from Eq.~\eqref{eq:vtrans_vs_q}] and $0.6v_{\text{trans}}$. They used $0.6v_{\text{trans}}$ instead of the obvious choice of $0.5v_{\text{trans}}$ because of the computational cost of using the lower velocity, especially for nearly equal-mass binaries, with the earlier, less efficient version of the code.
The results, summarized in Fig.~6 in~\cite{Johnson-McDaniel:2021rvv}, show that the differences in $\cos\theta_A^\infty$ for all $500$ systems are below $10^{-3}$, thus validating their choice of $v_{\rm trans}$. However, in our case, we use the natural choice of $0.5v_{\text{trans}}$, since we find that it is possible to evolve to $0.5v_{\text{trans}}$ in a reasonable amount of time, without incurring any hardware memory issues (which was the case in quasicircular code from~\cite{Johnson-McDaniel:2021rvv} for close to equal mass binaries). We also have a much larger sample set of $2000$ simulated BBHs to test our code, because we are sampling more input parameters, namely the orbital eccentricity as well as both $\psi_1$ and $\psi_2$ (instead of just $\phi_{12}$).


\subsection{Validating $v_{\rm trans}$}
\label{subsec:validating_vtrans}


To validate that the quasicircular $v_{\rm trans}$ indeed gives sufficient accuracy in the eccentric case, we now carry out the same convergence test performed in~\cite{Johnson-McDaniel:2021rvv} to validate $v_{\rm trans}$ [given by Eq.~(\ref{eq:vtrans_vs_q})], except that we are now using $v_{\rm trans}$ and $0.5v_{\rm trans}$ as our two transition velocities. We measure $\delta \cos \theta^{\infty}_{1}$ and $\delta \cos \theta^{\infty}_{2}$, which here denote the absolute values of the difference in cosines of spin tilt~1 and spin tilt~2 at infinity, respectively, obtained using $v_{\rm trans}$ and $0.5v_{\rm trans}$ as transition velocities (we use the same notation for other similar differences later).
However, as mentioned in the introduction, instead of demanding the same accuracy in the cosines of the tilts at infinity as in the quasicircular case, we instead only require an accuracy of better than $10^{-2}$, since the orbit-averaged eccentric equations we use are less accurate than the quasicircular ones ($2$PN in spinning terms and $3$PN in nonspinning terms in the eccentric code compared to $3$PN in spinning terms and $3.5$PN in nonspinning terms in the quasicircular code). In fact, we expect that we have errors of up to $\sim 0.1$ from the lack of $2.5$PN spin terms and $3.5$PN nonspinning terms, given the results in the quasicircular case (see Fig.~7 in~\cite{Johnson-McDaniel:2021rvv} for the spin terms and the discussion later for the nonspinning terms). Nevertheless, we keep $10^{-2}$ as our desired tolerance, since it is below the anticipated statistical errors for observations in the fifth observing run of the advanced GW detector network~\cite{Kulkarni:2023nes}.
Thus, along with the results in the quasicircular case, this suggests that once we have the $2.5$PN spin terms and $3.5$PN nonspinning terms in the orbit-averaged evolution equations, the $v_{\rm trans}$ we are using, or a scaling of it to slightly smaller values, would give an error of $< 10^{-2}$. \edit{Reference}~\cite{Johnson-McDaniel:2021rvv} also considered the difference in the tilts at infinity between the $2.5$PN and $3$PN spin terms and found that these are at most $\sim 10^{-2}$, so to ensure that the errors are $< 10^{-2}$, one will want to extend the orbit-averaged evolution to at least $3$PN in the spinning terms. As discussed in~\cite{Phukon:2025yva}, obtaining the orbit-averaged evolution at $3$PN only requires extending the quasi-Keplerian parameterization for arbitrary spins to $3$PN---all the other ingredients are already available in the literature.

\begin{figure*}
\includegraphics[width=.90\textwidth,height=6cm]{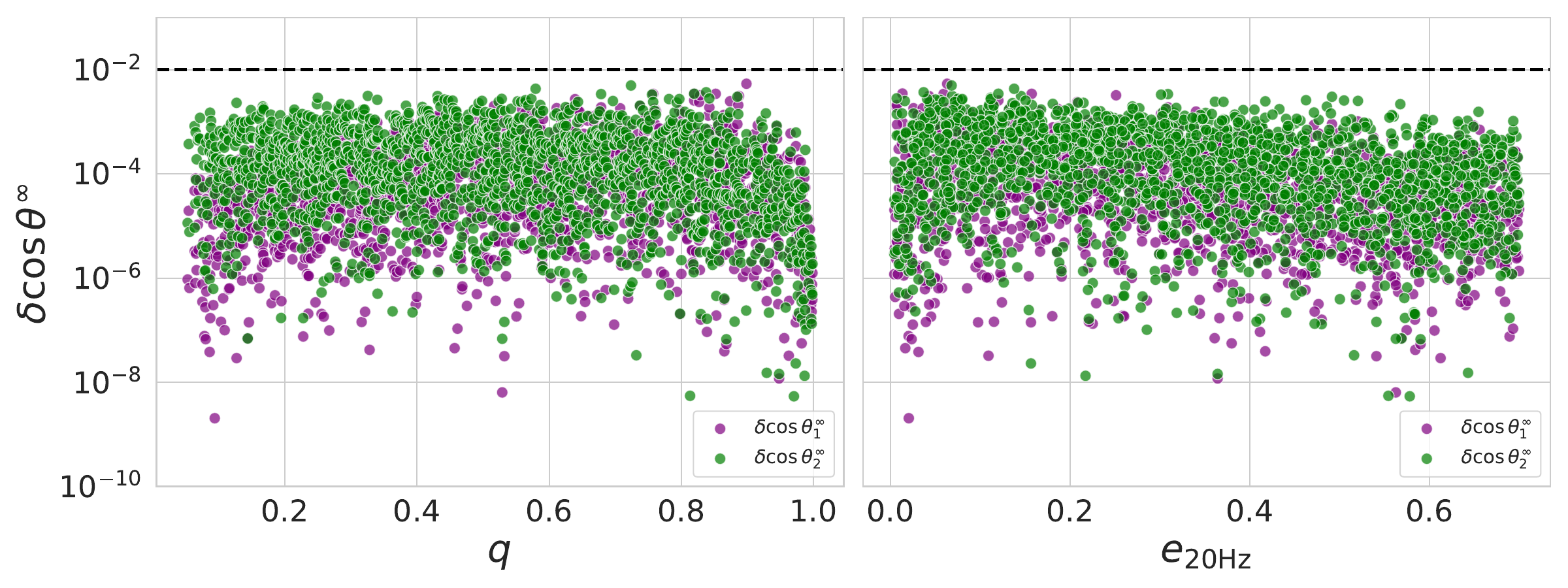}
\caption{Absolute values of the differences in cosines of spin tilts at infinity obtained using $v_{\rm trans}$ and $0.5v_{\rm trans}$ as the two transition orbital velocities. This plot restricts to the $1963$ binaries that reach $0.5v_{\rm trans}$ without the orbit-averaged evolution triggering a stopping condition. We plot these versus mass ratio on the left and eccentricity (at $20$~Hz) on the right. All the differences are well below our desired tolerance of $10^{-2}$, indicated by the dashed horizontal lines.}

\label{fig:delta_cos_tilt_inf_vs_q_and_ecc_and_etrans}
\end{figure*}

To validate $v_{\rm trans}$, we analyze a total of $2000$ binaries, with parameters spanning a broad range, only differing with respect to the distribution used in the quasicircular analysis~\cite{Johnson-McDaniel:2021rvv} in sampling the spin angles, where we sample the spin angles uniformly on the $2$-sphere, rather than uniformly in the tilt angles, as in the quasicircular case, and also sample both the $\psi_{1}$ and $\psi_{2}$ angles, since both enter into the orbit-averaged eccentric evolution, rather than just $\phi_{12}$, which is all that is necessary in the quasicircular case. 
The reference frequency \( f_{\text{ref}} \) is fixed at $20$~Hz, with initial eccentricities \( e_{\rm 20\, Hz} \) uniformly distributed between $0$ and $0.7$. We restrict the maximum eccentricity to $0.7$ for all the analyses in this paper because evolving backward from higher starting eccentricities triggered some of the checks that the PN equations we are using are not reliable at the initial conditions for many binaries. Specifically, the checks that commonly fail for $e_{\rm 20\, Hz} > 0.7$ are those checking that the binary is bound (i.e., that its energy is negative), that the binary's energy is increasing as the binary evolves backward (i.e., the time derivative of the binary's energy is negative), and that the PN expansion parameter $\bar{x}=v^{2}/(1-e^{2}) < 1$.

\begin{table}[h]
	\centering
	\caption{Parameters of binaries with maximum differences in $\cos\theta^\infty_1$ and $\cos\theta^\infty_2$ for the convergence test for eccentric binary evolution. Here $\delta \cos\theta^\infty_1$ and $\delta \cos\theta^\infty_2$ denote the absolute values of the differences in $\cos\theta^\infty_1$ and $\cos\theta^\infty_2$, respectively, obtained using $v_{\rm trans}$ and $0.5v_{\rm trans}$ as transition orbital velocity in hybrid evolution. The initial parameters of the binaries are specified at a reference frequency of $20~\rm{Hz}$. Binary 1 is the one exhibiting the largest $\delta \cos\theta^\infty_1$ and binary 2 is the one exhibiting the largest $\delta \cos\theta^\infty_2$. 
	\label{tab:convg_test_ecc_binaries}}
	\renewcommand{\arraystretch}{1.3}
	\setlength{\tabcolsep}{8pt}
	\begin{tabular}{lcc}
		\hline\hline
		Parameter & Binary 1 & Binary 2 \\
		\midrule
		$e_{\rm 20\,Hz}$ & 0.063 & 0.138 \\
		$m_1$, $m_2$ ($M_\odot$) & 17.49, 15.69 & 29.23, 16.94 \\
		$\chi_1$, $\chi_2$ & 0.081, 0.847 & 0.965, 0.580 \\
		$\theta_1$, $\theta_2$ & 1.852, 2.395 & 0.887, 1.654 \\
		$\psi_1$, $\psi_2$ & 4.460, 0.257 & 5.607, 2.697 \\
		$\delta\cos\theta^\infty_1$ & {$5.37\times10^{-3}$} & $5.77 \times10^{-4}$ \\
		$\delta\cos\theta^\infty_2$ & $1.47\times10^{-4}$ & $4.98\times10^{-3}$ \\
		\hline\hline
	\end{tabular}
\end{table}

\begin{figure}
	\includegraphics[width=.99\linewidth,height=6cm]{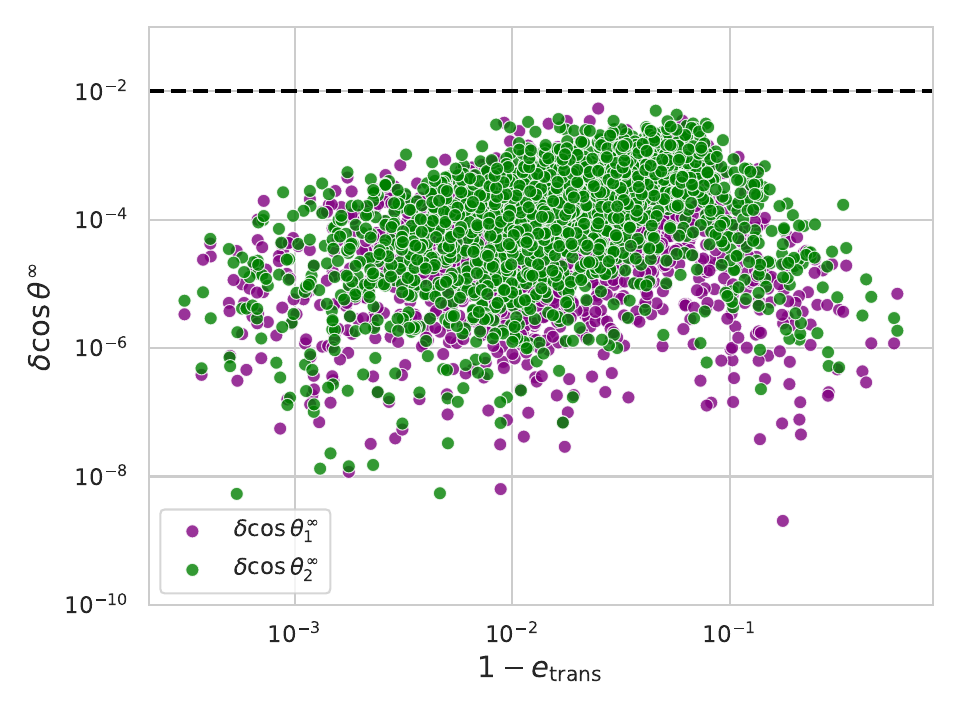}
	\caption{The absolute values of the differences in the cosines of spin tilts at infinity computed using the two transition velocities of $v_{\rm trans}$ and $0.5v_{\rm trans}$, versus the eccentricities at $v_{\rm trans}$, for the $1963$ binaries which successfully evolved to $0.5v_{\rm trans}$. Most of the eccentricities at transition are greater than $0.9$ and the distribution extends to above $0.999$. The dashed line has the same meaning as in Fig.~\ref{fig:delta_cos_tilt_inf_vs_q_and_ecc_and_etrans}.}
	\label{fig:delta_cos_tilts_vs_e_trans}
\end{figure}

In Fig.~\ref{fig:delta_cos_tilt_inf_vs_q_and_ecc_and_etrans}, we show the same convergence test as in the quasicircular case for the $1963$ binaries out of the $2000$ binaries for which the orbit-averaged evolution reaches a stopping frequency of $f_{\rm stop}$ which is within $1\%$ of the $f_{\rm trans}$ associated with $0.5v_{\rm trans}$ (corresponding to a $v_{\rm stop}$ within $0.33\%$ of the value of $0.5v_{\rm trans}$), without triggering any stopping condition.
The maximum differences in $\cos \theta^{\infty}_{1}$ and $\cos\theta^{\infty}_{2}$ are $\mathcal{O} (10^{-3})$. The binary parameters for both of these cases are given in Table~\ref{tab:convg_test_ecc_binaries}. We also plot the eccentricity at $v_{\rm trans}$ ($e_{\rm trans}$) in Fig.~\ref{fig:delta_cos_tilts_vs_e_trans}, where we find that most of them are greater than $0.9$, and some are even greater than $0.999$. Since the orbit-averaged approximation is less accurate for large eccentricities, future work will check the accuracy of the approximation by comparing the evolutions with and without orbit averaging.

We now discuss the binaries for which the orbit-averaged evolution stopped before reaching $v_{\rm trans}$ (binaries which reached $v_{\rm trans}$ successfully also reached $0.5 v_{\rm trans}$). This early termination of the evolution is because of different stopping conditions implemented in the orbit-averaged code to ensure that the equations we are evolving do not give obviously unphysical results at initial or any stage of the evolution.
Specifically, we consider the $11$ binaries whose GW frequency when the evolution stopped differs from the desired $f_{\rm trans}$ by more than $1\%$ (i.e., the final orbital velocity reached differs from the desired $v_{\rm trans}$ by more than $\sim 0.3\%$). Out of these $11$, $9$ of them had an initial eccentricity of $e_{\rm 20\, Hz}<0.004$, which is smaller than that allowed by the orbit-averaged evolution equations. This is because, due to the $2$PN spin-spin contributions to $\dot{e_t^2}$, eccentric, precessing binaries attain a minimum, nonzero value of $e_t$ when evolved forward. This causes $e_t^2$ to become negative when evolving backwards (see the discussion in~\cite{Klein:2010ti,Klein:2018ybm,Phukon:2025yva}). We find that the check that $\ddot{\omega}\le0$ gets triggered by the interpolated data points in these cases, but the integration is stopped because $e_t^2 < 0$. For the other $2$ binaries, the PN energy expression used in the evolution [Eq.~(C1a) in~\cite{Phukon:2025yva}] gave unbound binaries. On the other hand, all the $1963$ binaries which stopped within $1\%$ of $f_{\rm trans}$, stopped due to the evolution reaching the desired ending frequency.
Finally, $25$ binaries were not able to start their evolution because their initial conditions corresponded to a positive energy (i.e., the binary is not bound), and one binary did not evolve because the orbital angular momentum was not increasing when evolving backward in time (i.e., its time derivative is positive), at the initial conditions.

For binaries that stop their orbit-averaged evolution before reaching within $1\%$ of $f_{\rm trans}$, we start the precession-averaged evolution at the stopping frequency $f_{\rm stop}$. Since this is above the desired transition frequency, we expect that the tilts at infinity will be less accurate than if we had been able to evolve to the desired frequency $f_{\rm trans}$. To estimate how much accuracy we lose in such cases, we compare with the tilts at infinity obtained when transitioning to the precession-averaged evolution at $2v_{\rm stop}$. Since we are comparing with an earlier transition point, where the precession-averaged evolution will be less accurate, this gives an upper bound on the size of the errors. We notice that we are only able to evolve only $1$ out of $11$ binaries because most of them stopped very early in their evolutions (stopping frequencies between $11$ and $20$ Hz), thus a transition velocity of $2v_{\rm stop}$ would correspond to a frequency higher than $f_{\rm ref}=20$ Hz. The binary parameters for this case are listed in Table~\ref{tab:sole_binary_to_2vtrans}. We find that the difference in the cosine of tilt~1 is $2.5\times 10^{-3}$, while that of tilt~2 is $2.1\times 10^{-2}$. This shows that one still gets good accuracy from the hybrid evolution in most cases, even if the orbit-averaged evolution is not able to reach the desired transition frequency.

\begin{table}[h]
	\centering
	\caption{Parameters for the sole binary that successfully evolved to $2v_{\rm stop}$ out of 11 binaries whose GW frequency when the evolution stopped differed from the desired $f_{\rm trans}$ by more than $1\%$. Here $\delta\cos\theta^\infty_1$ and $\delta\cos\theta^\infty_2$ denote the absolute values of the differences in the cosines of tilts obtained using $v_{\rm stop}$ and $2v_{\rm stop}$ as transition orbital velocity. The initial eccentricity and spin angles are specified at $f_{\rm ref}= 20$ Hz. 
	\label{tab:sole_binary_to_2vtrans}}
	\renewcommand{\arraystretch}{1.3}
	\setlength{\tabcolsep}{8pt}
	\begin{tabular}{lc}
		\hline\hline
		Parameter & Value \\
		\midrule
		$e_{\rm 20\,Hz}$               & 0.004                 \\
		$m_1$, $m_2$ ($M_\odot$)        & 75.32, 63.33          \\
		$\chi_1$, $\chi_2$             & 0.959, 0.137          \\
		$\theta_1$, $\theta_2$         & 2.399, 0.499          \\
		$\psi_1$, $\psi_2$             & 0.898, 4.549          \\
		$\delta\cos\theta^\infty_1$  &  $2.5\times 10^{-3}$\\
		$\delta\cos\theta^\infty_2$ & $2.1\times 10^{-2}$ \\
		\hline\hline
	\end{tabular}
\end{table}

\begin{figure*}
	\includegraphics[width=.90\textwidth,height=6cm]{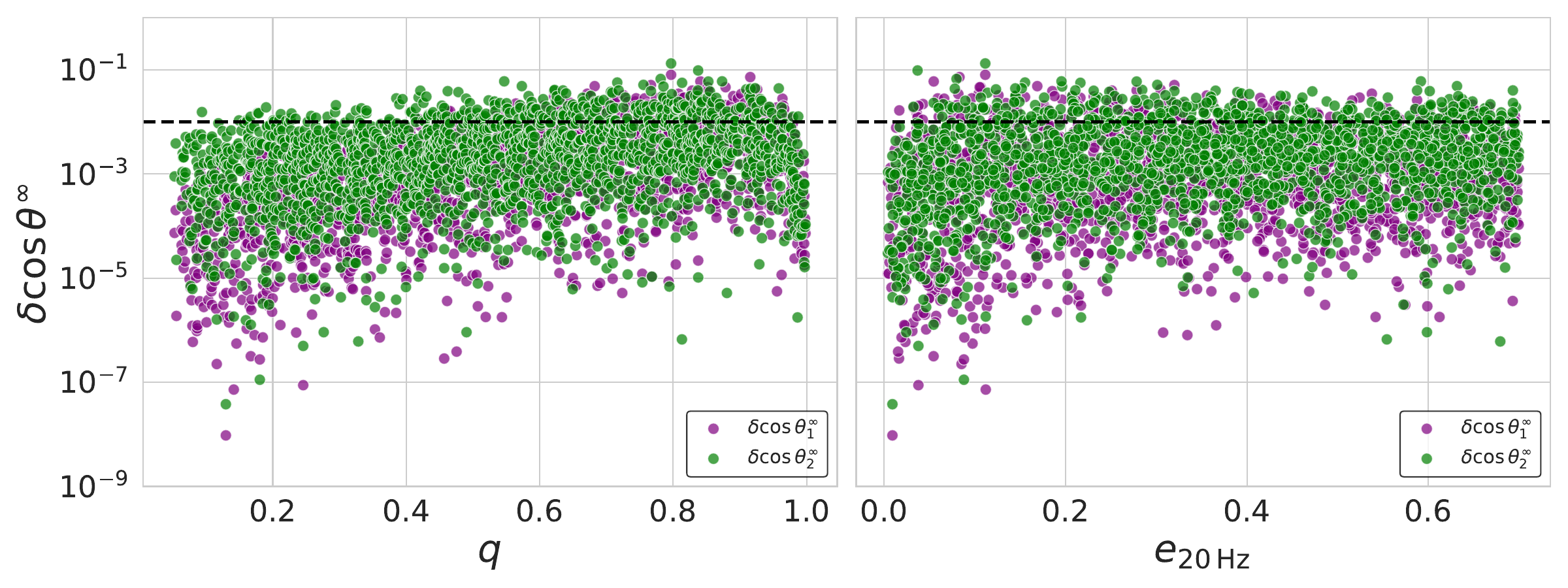}
	\caption{The absolute differences in the cosines of spin tilts at infinity computed using $2.5$PN and $3$PN nonspinning terms in the orbit-averaged evolution, versus mass ratio (left) and eccentricity at $20$~Hz (right) for the $1963$ binaries which successfully evolved to $v_{\rm trans}$ using both evolutions. The horizontal dashed line corresponds to $10^{-2}$. About $10\%$ of the binaries have differences greater than $10^{-2}$, though we expect that this overestimates the errors from the $3$PN truncation in the nonspinning terms used by default.}
	\label{fig:delta_cos_tilt_inf_bw_phaseO5_and_phaseO6}
\end{figure*}

\begin{table}[h]
	\centering
	\caption{Binary with both the maximum $\delta \cos\theta^\infty_1$ and maximum $\delta \cos\theta^\infty_2$ from comparing cosine tilts at infinity obtained using the 2.5PN and 3PN nonspinning contributions in the orbit-averaged evolution equations. The initial eccentricity and spin angles are specified at $f_{\rm ref}= 20$ Hz.
	\label{tab:2p5PN_vs_3PN_non_spin}}
	\renewcommand{\arraystretch}{1.3}
	\setlength{\tabcolsep}{8pt}
	\begin{tabular}{lc}
		\hline\hline
		Parameter & Value \\
		\midrule
		$e_{\rm 20\,Hz}$ & 0.112 \\
		$m_1$, $m_2$ ($M_\odot$) & 99.10, 78.97 \\
		$\chi_1$, $\chi_2$ & 0.857, 0.651 \\
		$\theta_1$, $\theta_2$ & 0.564, 2.526 \\
		$\psi_1$, $\psi_2$ & 6.143, 5.158 \\
		$\delta\cos\theta_1^\infty$ & $7.9 \times 10^{-2}$ \\
		$\delta\cos\theta_2^\infty$ & $1.3 \times 10^{-1}$ \\
		\hline\hline
	\end{tabular}
\end{table}

We also compare the differences we get in the cosines of the tilts at infinity obtained using the $2.5$PN and $3$PN nonspinning terms in the orbit-averaged evolution in Fig.~\ref{fig:delta_cos_tilt_inf_bw_phaseO5_and_phaseO6}. Out of the total $1963$ binaries which successfully evolved to $v_{\rm trans}$ with both orders of the nonspinning terms, $167$ binaries have $\delta\cos\theta^{\infty}_{1}>10^{-2}$, while $294$ binaries have $\delta\cos\theta^{\infty}_{2}>10^{-2}$. 
We find the same binary produced both the maximum $\delta\cos\theta_1^\infty$ and $\delta\cos\theta_2^\infty$ values, which are $7.9 \times 10^{-2}$ and $1.3 \times 10^{-1}$, respectively. The parameters of this binary are listed in Table~\ref{tab:2p5PN_vs_3PN_non_spin}. 

Since~\cite{Johnson-McDaniel:2021rvv} did not check the differences due to the PN order of the nonspinning terms, we perform this analysis for quasicircular binaries using the default PN approximant (TaylorT5) in the orbit-averaged evolution.\footnote{The TaylorTX approximant is referred to as \texttt{SpinTaylorTX} in the quasicircular code, for consistency with the waveform models in LALSuite.} (This uses the same $1963$ binaries used in the eccentric case with the eccentricity set to zero.) Here the maximum absolute differences in the cosines of the tilts at infinity estimated using $3$PN and $3.5$PN accurate orbit-averaged evolution (the highest two orders available in the quasicircular code) are $\delta\cos\theta^{\infty}_{1}=3.5\times 10^{-3}$ and $\delta\cos\theta^{\infty}_{2}=3.3\times 10^{-3}$. These are somewhat above the desired tolerance of $10^{-3}$ for the quasicircular code, but we expect that these differences provide an upper bound on the errors due to using the $3.5$PN evolution equations, which are likely at most around $10^{-3}$, since the maximum differences when using the $2.5$PN and $3$PN nonspinning terms are a factor of $\sim 2$ larger ($8.4\times 10^{-3}$ and $7.8\times 10^{-3}$, respectively). We leave an explicit comparison with the $4$PN and $4.5$PN nonspinning terms from~\cite{Blanchet:2023bwj} for future work. To enable a direct comparison of the eccentric evolution (which uses TaylorT4) with the quasicircular evolution, we compute the differences in cosine tilts using the $2.5$PN and $3$PN nonspinning terms for the quasicircular evolution using the TaylorT4 approximant. We find that the maximum values of $\delta\cos\theta^{\infty}_{1}$ and $\delta\cos\theta^{\infty}_{2}$ are $2.3\times 10^{-2}$ and $2.5\times 10^{-2}$ respectively. Thus, since the maximum differences in cosine tilts at infinity using $3$PN and $3.5$PN terms with TaylorT4 in quasicircular evolution are $6.7\times 10^{-3}$ and $3.3\times 10^{-3}$, respectively, it is likely that the errors due to the truncation at $3$PN for the nonspinning terms in the eccentric case are a few times $10^{-2}$. (This is obtained by computing the ratio of the absolute differences in cosine tilts between $3$PN and $3.5$PN to those between $2.5$PN and $3$PN in the quasicircular case and then multiplying by the differences in cosine tilts between $2.5$PN and $3$PN in the eccentric case.)

\subsection{Analysis for very high eccentricities}

Apart from the results in Fig.~\ref{fig:delta_cos_tilt_inf_vs_q_and_ecc_and_etrans} where initial eccentricity was varied from $0$ to $0.7$, we also evolved $1000$ highly eccentric binaries, where $e_{\rm 20\, Hz}$ ranged from $0.7$ to $0.99$, and all other parameters are sampled from the ranges defined in earlier sections. Only $166$ of these binaries successfully completed their evolution. The other $834$ of them were not able to start their evolution because one of the internal consistency checks of the PN expressions failed for the initial conditions. For most of the binaries ($697$), the problem was that the PN energy expression used in the evolution [Eq.~(C1a) in~\cite{Phukon:2025yva}] gave an unbound binary, while many others ($134$ binaries) had $\bar{x} \ge 1$ initially, so we do not trust the PN equations we use. For one of the binaries, the initial value of the energy time derivative is positive (while it should be negative due to GW emission), while for the other two binaries this time derivative becomes positive during the evolution. For the $166$ binaries which evolved successfully to $v_{\rm trans}$, they were successfully able to evolve to $0.5v_{\rm trans}$ and we found that the differences are all lower than our desired $10^{-2}$ tolerance.

\subsection{Runtime}
\begin{figure*}
	\centering
		\includegraphics[width=.90\textwidth]{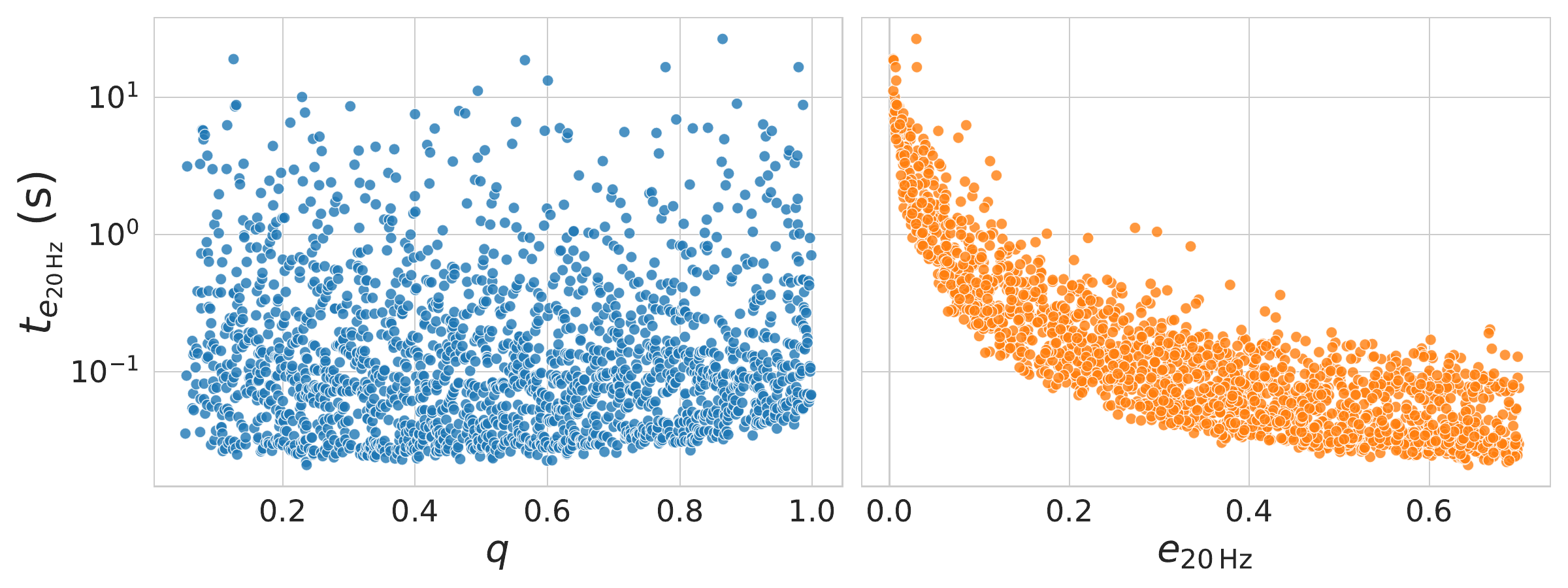}

	\caption{The time taken by the eccentric hybrid evolution code to compute the spin tilts at infinity for each of $1963$ binaries which successfully evolved to $v_{\rm trans}$. We show the variation along with mass ratio $q$ (left) and starting eccentricity $e_{\rm 20\, Hz}$ (right). There is a strong correlation observed between the runtime and $e_{\rm 20\, Hz}$, as binaries with larger starting eccentricities evolve more quickly (in physical time).}
	
	\label{fig:ecc_runtime_comparison}
\end{figure*}

For our $1963$ successful binaries, we calculated the time it takes to compute the spin tilts at infinity using our hybrid evolution code, averaging over $5$ evolutions for each binary. These tests were performed on a heterogeneous computing pool with a variety of CPUs, including Intel's Skylake-X (2017) and more recent models, as well as AMD's Zen~4 (2022) and later generations, with base clock speeds ranging between 2.2~GHz and 4.7~GHz. The results are presented in Fig.~\ref{fig:ecc_runtime_comparison}.

The eccentric hybrid evolution code runtimes are roughly in the same range as those obtained using the quasicircular hybrid evolution code (see Fig.~10 in~\cite{Johnson-McDaniel:2021rvv}), though they are smaller for larger eccentricities, where the binary's physical evolution is quicker. This is why as the mass ratio approaches $1$, the eccentric code becomes faster than the quasicircular code for the vast majority of BBHs which we consider.

\section{Comparison with other types of evolutions}
\label{sec:spin_tilts_inf}

\subsection{Comparison with quasicircular evolution}

\begin{table}[h]
	\centering
	\caption{Binaries with maximum $\delta \cos\theta^\infty_1$ and $\delta \cos\theta^\infty_2$ from the comparison of eccentric and quasicircular hybrid evolutions. Here $\delta \cos\theta^\infty_1$ and $\delta \cos\theta^\infty_2$ denote the absolute values of the differences in cosines of tilts at infinity obtained using the eccentric hybrid evolution and the quasicircular hybrid evolution (taking the eccentricity to be zero). Binary 1 is the one exhibiting the largest $\delta \cos\theta^\infty_1$ and binary 2 is the one exhibiting the largest $\delta \cos\theta^\infty_2$. The initial parameter values of the binaries are specified at $f_{\rm ref}= 20$ Hz.
	\label{tab:qc_vs_ecc_hyb_evol}}
	\renewcommand{\arraystretch}{1.3}
	\setlength{\tabcolsep}{8pt}
	\begin{tabular}{lcc}
		\hline\hline
		Parameter & Binary 1 & Binary 2 \\
		\midrule
		$e_{\rm 20\,Hz}$ & 0.512 & 0.476 \\
		$m_1$, $m_2$ ($M_\odot$) & 95.90, 77.72 & 48.39, 45.78 \\
		$\chi_1$, $\chi_2$ & 0.212, 0.972 & 0.289, 0.003 \\
		$\theta_1$, $\theta_2$ & 0.161, 2.535 & 0.067, 2.834 \\
		$\psi_1$, $\psi_2$ & 0.519, 4.912 & 4.956, 6.219 \\
		$\delta\cos\theta^\infty_1$ & $5.97\times10^{-1}$ & $1.61\times10^{-1}$ \\
		$\delta\cos\theta^\infty_2$ & $1.57\times10^{-2}$ & $1.75$ \\
		\hline\hline
	\end{tabular}
\end{table}

\begin{figure*}
	\includegraphics[width=.90\textwidth,height=6cm]{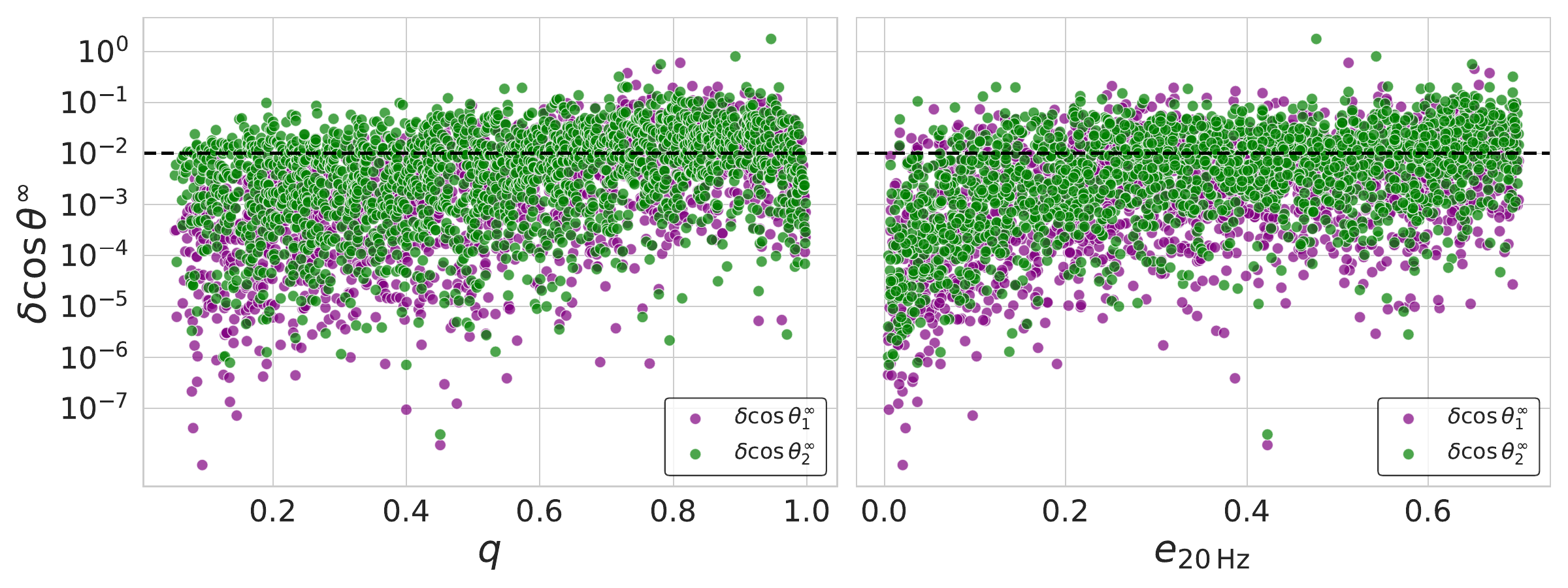}
	\caption{The absolute values of the differences in the cosines of the spin tilts at infinity obtained using the eccentric hybrid evolution and the quasicircular evolution (taking the eccentricity to be zero and using the same PN orders as in the eccentric hybrid evolution), for the $1963$ binaries which successfully evolved to $v_{\rm trans}$.}
	\label{fig:delta_cos_tilt_inf_vs_q_and_e0_same_settings}
\end{figure*}

Here we illustrate that there are portions of parameter space where eccentricity has a significant effect on the tilts at infinity. We do this by comparing the results of the eccentric hybrid evolution with the quasicircular hybrid evolution (from~\cite{Johnson-McDaniel:2021rvv}) applied to the set of the $1963$ binaries which successfully evolved to $v_{\rm trans}$ (where we just ignore the eccentricity when applying the quasicircular code). We use the same PN orders ($3$PN nonspinning and $2$PN spinning) for the quasicircular evolution as are used in the eccentric evolution, as well as the same approximant (TaylorT4), for an exact comparison. We show the differences in Fig.~\ref{fig:delta_cos_tilt_inf_vs_q_and_e0_same_settings}, where we find that $476$ binaries have $\delta \cos \theta^{\infty}_{1} \ge 10^{-2}$, and $703$ binaries have $\delta \cos \theta^{\infty}_{2} \ge 10^{-2}$. These differences are similar to or greater than unity in a few cases. The maximum absolute difference obtained for the cosine of tilt~1 was $\sim 0.6$, while the maximum cosine tilt~2 difference was $\sim 1.8$. The binary parameters of these two cases are listed in Table~\ref{tab:qc_vs_ecc_hyb_evol}.

\subsection{Hybrid evolution vs precession-averaged evolution}

\begin{figure*}
	\includegraphics[width=.90\textwidth,height=6cm]{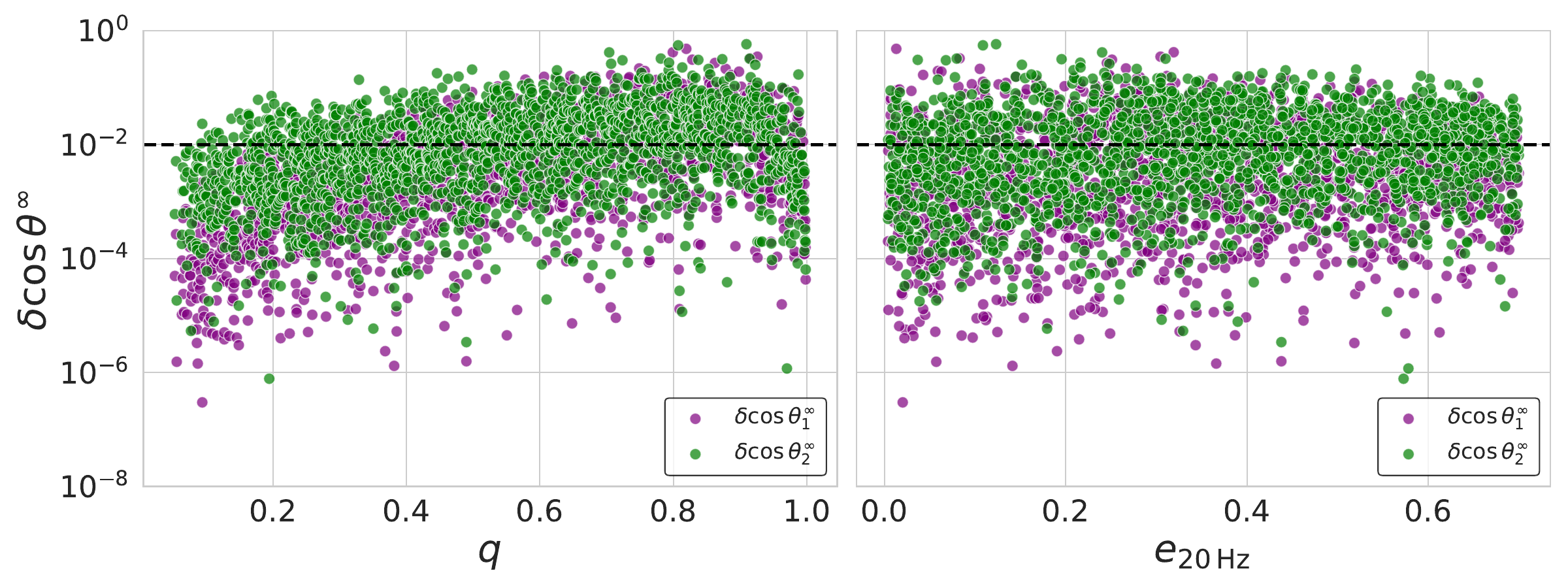}
	\caption{The absolute values of the differences between the cosines of spin tilts at infinity computed using the hybrid evolution code and using only precession-averaged evolution, for the $1963$ binaries which successfully evolved to $v_{\rm trans}$. Some of the differences are much higher than $10^{-2}$, and even close to $1$ in some cases, showing the importance of using the hybrid evolution.}
	\label{fig:delta_cos_tilt_inf_vs_q_and_e0_hyb_vs_prec_only}
\end{figure*}

We now compare the tilts at infinity estimates from the only precession-averaged evolution and from the full hybrid evolution. We consider the same $1963$ binaries which successfully evolved to $v_{\rm trans}$ from the $2000$ binaries described earlier in Sec.~\ref{subsec:validating_vtrans}, and show the results in Fig.~\ref{fig:delta_cos_tilt_inf_vs_q_and_e0_hyb_vs_prec_only}.
We observe that there are large differences ($\ge 10^{-1}$ and some approaching unity) in the spin tilts at infinity for a significant number of binaries, particularly for more symmetric mass binaries ($q$ close to $1$) and lower values of starting eccentricities. These maximum differences are significantly larger than those found for the quasicircular evolution, where the differences are $\lesssim 10^{-1}$ (see Fig.~8 in~\cite{Johnson-McDaniel:2021rvv}). The comparison in~\cite{Johnson-McDaniel:2021rvv} uses a higher-order PN expression to compute the initial orbital angular momentum for the only precession-averaged evolution than the $1$PN one used in the hybrid evolution, since this is found to give slightly better agreement between the two evolutions, while we just use the Newtonian expression used in the hybrid evolution here. However, the differences between the two evolutions in the quasicircular case are still only $\lesssim 10^{-1}$ when using the $1$PN expression for the orbital angular momentum for the only precession-averaged evolution.

\section{Tilts at infinity vs tilts at $\lowercase{e} \nearrow 1$ and at formation}

\label{sec:inf_vs_form}

\begin{figure}
	\centering
		\centering
		\includegraphics[width=\linewidth]{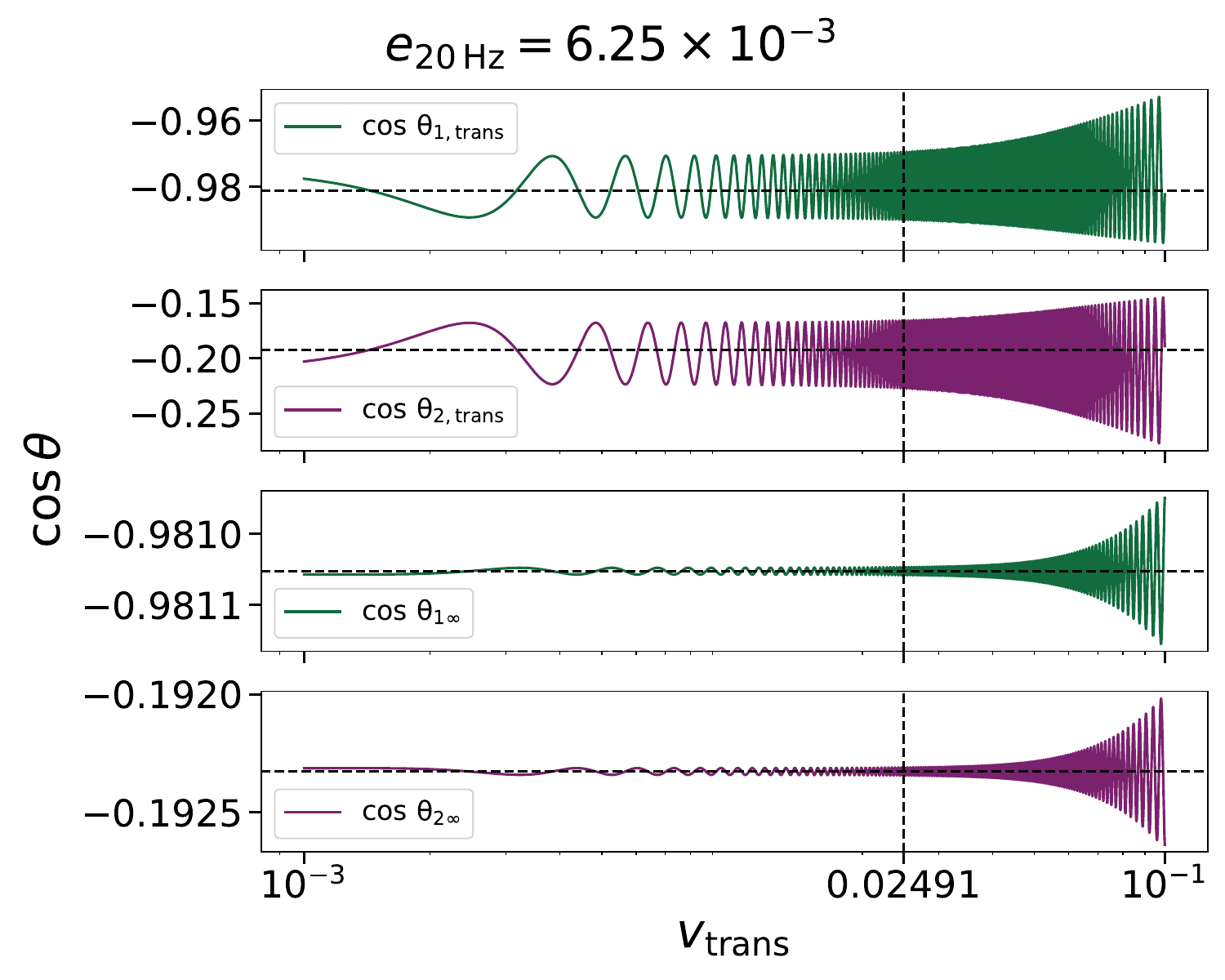}

		\centering
		\includegraphics[width=\linewidth]{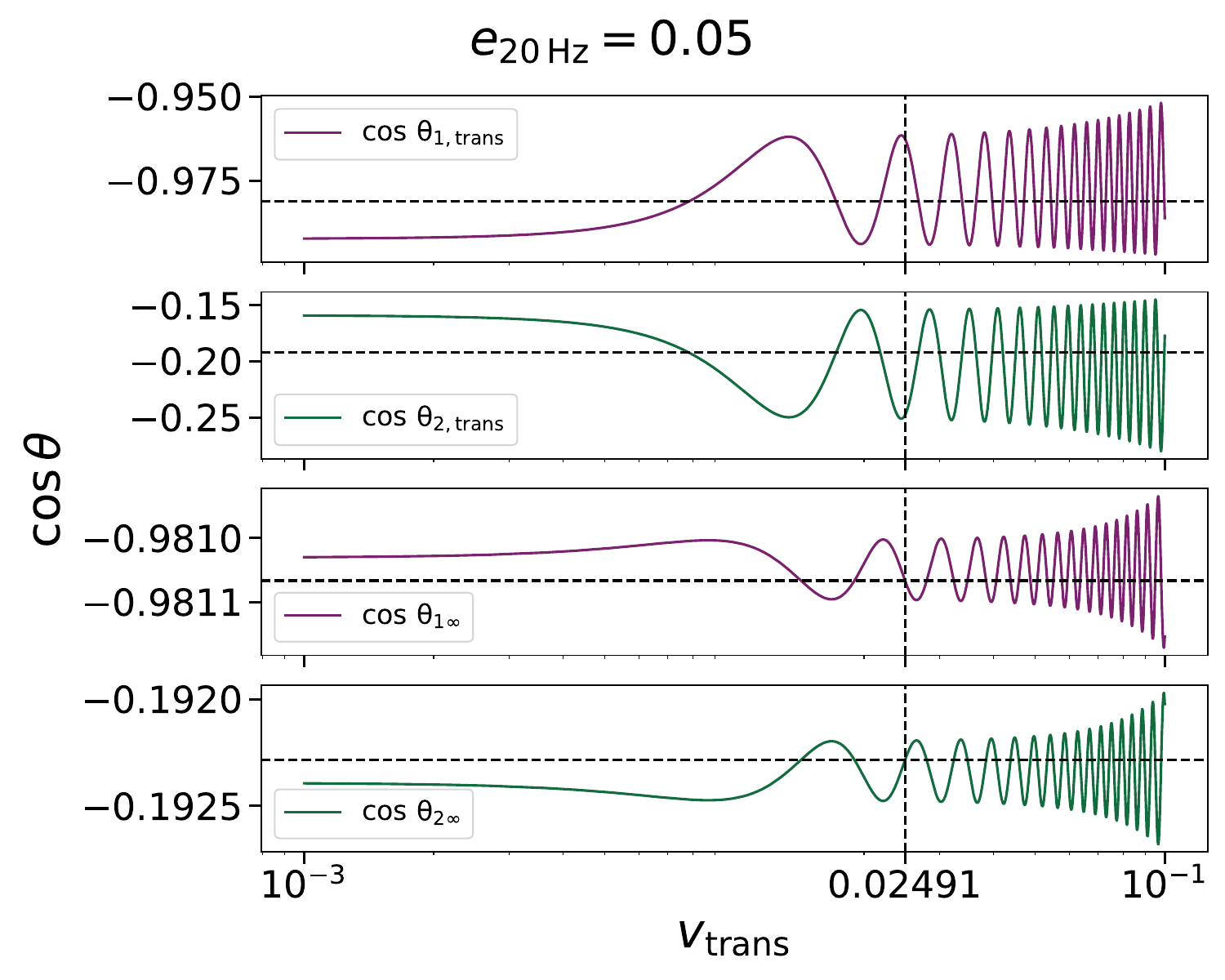}

	\caption{Variation of spin tilts at transition orbital velocity $v_{\rm trans}$ and at infinity, with $v_{\rm trans}$, for binaries with two different starting eccentricities (binary A in the upper subfigure and binary B in the lower subfigure). All the other binary parameters are fixed to the values given in Table~\ref{tab:fig_7A_binary}. The upper panels in each subfigure show the cosines of tilts at different $v_{\rm trans}$ values, while the lower panels show the cosines of spin tilts at infinity. The vertical dashed black line denotes the empirical value of the $v_{\rm trans}$ [from Eq.~\eqref{eq:vtrans_vs_q}], while the horizontal dashed black line denotes the tilts at infinity computed using that empirical $v_{\rm trans}$.}
	\label{fig:cosine_tilts_vs_vtrans}
\end{figure}

As discussed in Sec.~\ref{sec:intro}, the $L$ for eccentric binaries is bounded from above, so there is necessarily a lower bound on the \edit{intrinsic} uncertainty in the tilts at formation for eccentric binaries \edit{due to the unknown precessional phase}, unlike the quasicircular case where this uncertainty can in principle be arbitrarily small. Nevertheless, the maximum $L$ value attained for eccentric binaries (as $e \nearrow 1$) can still be large enough that there is minimal \edit{intrinsic} uncertainty in the tilts at formation. However, in other cases, particularly those with larger eccentricities at the reference frequency, there is a non-negligible uncertainty in the tilts at formation, and the tilts one obtains when evolving back to $e$ values close to $1$ with only orbit-averaged evolution will be a better approximation to the tilts at $e \nearrow 1$ than the tilts at infinity. \edit{Of course, the formation state of a BBH is not the same as the $e \nearrow 1$ limit, which corresponds to an infinite separation, but a finite $L$, whereas the tilts at infinity are obtained in the limit where $L \to \infty$. For instance, binaries can be dynamically assembled at a finite separation with very high eccentricities ($e \simeq 0.9999$). However, the tilts at $e \nearrow 1$ can be closer to those at formation than the tilts at infinity are, since  evolving the binary backward to $e \nearrow 1$ will not produce significant differences between the tilts at $e \nearrow 1$ and those at formation.}

\edit{We illustrate this by comparing the tilts at $e \nearrow 1$ and at formation for binaries from the Rapster~\cite{Kritos:2022ggc} open data~\cite{konstantinos_2024_10632717} with $e \ge 0.9999$ at formation, giving us a median value of $L/M^{2}=1.99$ and a minimum value of $L/M^{2}=0.83$ at formation. Here we compute the tilts at $e \nearrow 1$ using the method discussed below, specifying a tolerance of $10^{-3}$ in the cosines of the tilts. For the $55$ BBHs of these for which we were able to obtain the tilts at $e \nearrow 1$ in a computationally efficient manner, $50$ of them show cosine tilt differences lower than $10^{-3}$ between formation and at $e \nearrow 1$, and the other five have cosine tilt differences below $10^{-2}$. For each of these five binaries, the orbital velocity at formation is larger than the velocity to which the binary must be evolved backward for the bound on the remaining change in the cosine tilts up to $e \nearrow 1$ to fall below our chosen tolerance of $10^{-3}$. Thus, it is not surprising that the differences between the tilts at $e \nearrow 1$ and those at formation for some such cases are higher than the tolerance, which applies only to the evolution from that stopping point to $e \nearrow 1$. In summary, these results show that for highly eccentric binaries, the $e \nearrow 1$ limit can work very well as an approximation to the formation of the BBH.}

\begin{table}[h]
	\centering
	\caption{Parameters of the binary at 20 Hz chosen for the example demonstrated in Fig.~\ref{fig:cosine_tilts_vs_vtrans}. Only the starting eccentricities differ ($e_{\rm 20 Hz}=6.25\times 10^{-3}$ for binary A and $e_{\rm 20 Hz}=0.05$ for binary B).
	\label{tab:fig_7A_binary}}
	\renewcommand{\arraystretch}{1.3}
	\setlength{\tabcolsep}{8pt}
	\begin{tabular}{lc}
		\hline
		\hline
		Parameter & Value \\
		\midrule
		$m_1$, $m_2$ ($M_\odot$)     & 7.67, 6.42    \\
		$\chi_1$, $\chi_2$          & 0.88, 0.35    \\
		$\theta_1$, $\theta_2$    & 2.93, 1.77    \\
		$\psi_1$, $\psi_2$       & 0, 4.25       \\
		\hline
		\hline
	\end{tabular}
\end{table}

\edit{We now} give an example of an eccentric case where the tilts at infinity are not a particularly good approximation to the tilts at $e \nearrow 1$ (differences above $10^{-2}$) as well as a case where they are a better approximation. Here, we use the tilts in the limit $e\nearrow 1$ as a proxy for the tilts at the binary's formation, since we are not considering any particular astrophysical formation scenario. However, we observe that the dynamically formed binaries tend to have large eccentricities at formation. For example, among the dynamical binaries considered in Appendix~\ref{app:pop_syn_evol}, the maximum and median eccentricities \edit{at formation} are $0.999998$ and $0.99$ respectively. We also show how the tilts at infinity depend on the transition velocity as well as how the tilts at different transition points vary, for comparison with the same illustration in the quasicircular case in Fig.~5 of~\cite{Johnson-McDaniel:2021rvv}.
Specifically, Fig.~\ref{fig:cosine_tilts_vs_vtrans} shows the behavior of the same BBH with two different starting eccentricities, $e_{\rm 20\, Hz}=6.25 \times 10^{-3}$ (binary~A) and $e_{\rm 20\, Hz}=0.05$ (binary~B).
We chose the binary parameters quasi-randomly for this illustration, and their values are listed in Table~\ref{tab:fig_7A_binary} at a reference frequency of $20$~Hz. We choose two starting eccentricities $e_{\rm 20\, Hz}=6.25 \times 10^{-3}$ and $0.05$, to contrast the different values of $L/M^2$ attainable as $e \nearrow 1$. We compute the $L/M^2$ values as $e \nearrow 1$ (denoted as $L_{e \nearrow 1}/M^2$) by evolving the binary backwards from $f_{\rm ref}$ to $v_{\rm trans}$ using orbit-averaged evolution and then using the leading-order expression from Eq.~(5.11) in Peters~\cite{PhysRev.136.B1224} to compute the value in the limit $e \nearrow 1$; we have checked that evolving to $v_{\rm trans}/2$ instead and then applying the Peters expression produces negligible differences. We find $L_{e \nearrow 1}/M^2$ to be $7.9$ and $4.1$ for binaries~A and~B, respectively. We thus expect the tilts at $e \nearrow 1$ for binary~A to have smaller intrinsic uncertainties than those for binary~B, since \edit{these} uncertainties in the tilts at a finite $L$ go as $M^2/L$ in the precession-averaged approximation~\cite{Gerosa:2015tea}. As an illustration, we take the spin angles at the smallest value of $v_\text{trans} = 10^{-3}$ plotted in Fig.~\ref{fig:cosine_tilts_vs_vtrans} and the $L_{e \nearrow 1}/M^2$ values, and use Eqs.~(23) in~\cite{Johnson-McDaniel:2021rvv} to obtain differences in the cosines of tilt~1 and tilt~2 over a precessional cycle of $0.02$, $0.06$ for binary~A and $0.03$, $0.09$ for binary~B.

\edit{We} see a notable difference in the behavior of the tilts at small values of $v_\text{trans}$ in Fig.~\ref{fig:cosine_tilts_vs_vtrans}. While in both cases, the tilts at infinity approach a constant value, for binary~A, the tilts at small values of $v_\text{trans}$ still seem to be oscillating around the tilts at infinity, while for binary~B they have apparently asymptoted to constants different from the tilts at infinity (differences of $0.01$ and $0.03$ for tilts~1 and~2, respectively). We can show that there is no significant evolution of the tilts for binary~B between \edit{$v = 10^{-3}$} and $v=0$ using the bounds derived in Appendix~\ref{app:delta_theta_bound}, where the eccentricity of binary~B is $0.999863$ at $v_\text{trans} = 10^{-3}$. The maximum absolute difference in cosine tilt~1 and~2 values \edit{between $v=10^{-3}$ and $v = 0$} is $1.3\times 10^{-3}$ and $4.0 \times 10^{-3}$. These are more than $6$ times smaller than the differences \edit{between the tilts at infinity and at $v=10^{-3}$}. However, for the eccentricity of binary~A of $0.999490$ at \edit{$v = 10^{-3}$}, those upper bounds are $0.018$ and $0.054$,\footnote{Likely not coincidentally, the upper bounds in this case are quite similar to the differences in the cosines of the tilts over a precessional cycle for this binary.} which are larger than the tilt differences between $v_\text{trans} = 10^{-3}$ and at infinity ($4\times10^{-3}$ and $0.01$, respectively, for tilts~1 and~2). Future work will consider if these conclusions change significantly when evolving without orbit averaging. If there is not a significant change to these conclusions, then we will also determine where in the eccentric parameter space it is better to compute the tilts at infinity as a proxy for the tilts at formation (as for binary~A) and where it is better to evolve backwards using just the orbit-averaged evolution to estimate the tilts at formation (as for binary~B).

We also compare the tilts at infinity with those at formation for a subset of the dynamically formed BBHs from~\cite{Kritos:2022ggc}. Specifically, we consider the subset of binaries from the data release~\cite{konstantinos_2024_10632717} where both black holes' spins are nonzero, as discussed in Appendix~\ref{app:pop_syn_evol}. At formation, these binaries have $L/M^2$ values ranging from $0.4$ to $140.7$. We evolve those binaries forward from their formation to $10$~Hz using our orbit-averaged eccentric code. For the $660$ binaries which evolve successfully without being computationally very expensive (discussed in Appendix~\ref{app:pop_syn_evol}), we then evolve back from $10$~Hz to infinity using our hybrid evolution code, and compute the corresponding spin tilts.\footnote{In this backward evolution, it is necessary to initialize the periastron advance per orbit $k$ using the result of the forward evolution rather than using the default computation in the orbit-averaged code, as discussed in Sec.~III~C in~\cite{Phukon:2025yva}, since there are certain cases where the evolution is fine-tuned enough that if one uses the default value of $k$ to initialize the backwards evolution, the $e^{2}_{t}<0$ stopping condition will be triggered. Improving the initialization of $k$ in the orbit-averaged code is thus an important avenue for future work.} These $660$ binaries have $L/M^2$ values ranging from $0.8$ to $104.2$ \edit{at formation}.

We find that, for a subset of systems, the absolute differences between the cosines of the spin tilts at formation and at $L \to \infty$ are small. In particular, $24\%$ ($18\%$) of binaries have differences in the cosine of tilt 1 (2) less than $10^{-2}$, and in some cases the differences are as small as $10^{-4}$. For the systems with both differences lower than $10^{-2}$, the values of $L/M^2$ at formation span the range $0.86$ to $24.33$. We have checked that the systems that have small differences between the tilts at formation and at $L \to \infty$ are not just the systems for which the tilts do not change significantly between formation and $10$~Hz: At most $10\%$ of the systems with the cosines of both of the tilt differences between formation and $L\to\infty$ less than $10^{-2}$ have similarly small differences in those quantities between formation and $10$~Hz. However, there are also cases where the tilts at $L \to \infty$ do not agree well with the tilts at formation, with differences between in their cosines as large as $\mathcal{O}(1)$ in $6\%$ of binaries. The binaries with larger differences have larger eccentricity at formation: The minimum eccentricity at formation is $0.9745$ for absolute differences larger than $0.5$. We compare these differences with the intrinsic uncertainties in the cosines of the spin tilts at formation arising from the finite precessional amplitude (see, e.g.,~\cite{Johnson-McDaniel:2021rvv}), i.e., the uncertainties if one does not know exactly when in a precessional cycle the binary formed, which is the case if one is only evolving with precession-averaged evolution. In this comparison, we find that approximately $52\%$ of the binaries exhibit differences that are smaller than these intrinsic uncertainties, which themselves are below $0.1$ for $73\%$ of the cases.

We also consider whether there are any of these binaries for which the tilts at $e \nearrow 1$ are a better approximation to the tilts at formation than the tilts at $L \to \infty$. We thus use the upper bound on the differences in the cosines of the spin tilts between formation and the $e \nearrow 1$ limit derived in Appendix~\ref{app:delta_theta_bound} to determine the point at which the cosines of the binary's spin tilts would both differ by less than $10^{-3}$ from their values at $e \nearrow 1$ and evolve the binaries back to that point using the orbit-averaged evolution code. We find that in $30\%$ of the cases, the spin tilts at $e \nearrow 1$ are closer to their values at formation than are the tilts at infinity, i.e., the behavior shown in the lower subfigure of Fig.~\ref{fig:cosine_tilts_vs_vtrans}, with cosine differences again below $10^{-2}$ for $16\%$ of the cases. Again, these cases are not just the ones where the tilts do not evolve significantly from formation to $10$~Hz, with less than $2\%$ of the binaries that have differences smaller than $10^{-2}$ between formation and at $e \nearrow 1$ having a similarly small difference between formation and $10$~Hz. 

\edit{Thus, this} study demonstrates that for eccentric binaries, the orbital angular momentum at formation can indeed be large enough that the spin tilts at infinity (i.e., $L \to \infty$) provide a good approximation to the spin tilts at formation. However, there are also cases where the tilts at $e \nearrow 1$ are a better approximation to the tilts at formation. We leave determining the regions of parameter space where it is appropriate to quote the tilts at $e \nearrow 1$ in preference to the tilts at infinity to future work.

\section{Summary and Future Directions}
\label{sec:concl}

This paper presents the first public hybrid evolution code for eccentric, precessing BBHs to compute tilts at infinity and studies its accuracy. The hybrid evolution approach we employ uses both orbit-averaging and precession-averaging at different orbital separations, transitioning between the two at an empirically determined transition orbital velocity $v_{\rm trans}$, to evolve BBHs from a reference frequency backwards in time to compute tilts at infinity. We have shown that there are significant differences (up to order unity) in the cosines of spin tilts obtained using our hybrid evolution compared to the only-precession-averaged evolution that has been used in~\cite{Fumagalli:2024gko}, which already showed that neglecting a small but nonzero residual eccentricity in the detectors' band ($e_{\rm 10 Hz} \le 0.05$), can lead to significant biases once the binary is evolved backwards to a large separation ($\sim 10^{4}M$). Our code is public~\cite{amiteshgit}, and it will be part of LALSuite~\cite{LALSuite} after being reviewed, so it can be used for such studies. We also have released the data generated by this code which is used for creating all the figures \edit{in}~\cite{singh_et_al_2025_zenodo}. Additionally, our code can be used to compute the tilts at infinity when analyzing compact binary GW signals using an eccentric, precessing BBH waveform model (which have started to be developed~\cite{Liu:2023ldr,Gamba:2024cvy}), though some work will be necessary to convert the eccentricity definition used in the cited waveform models to the eccentricity definition used in our code.

With the differences in cosines of spin tilts at infinity lower than $\sim 10^{-2}$, we have shown that the transition orbital velocity $v_{\rm trans}$ used in the quasicircular hybrid evolution in~\cite{Johnson-McDaniel:2021rvv} produces acceptable accuracy in the tilts at infinity for eccentric cases. These differences are lower than the anticipated statistical \edit{measurement} uncertainties in the fifth observing run of the advanced GW detector network. However, given the checks we made of the accuracy of this code and the quasicircular one, we also expect that it will be necessary to extend the eccentric orbit-averaged evolution from its current $2$PN accuracy in spinning terms and $3$PN accuracy in nonspinning terms to $3$PN and $3.5$PN, respectively, to obtain an overall accuracy of $< 10^{-2}$. Additionally, it will be necessary to check the accuracy of our orbit-averaged evolution against an evolution without orbit averaging, particularly since the backwards evolution leads to large eccentricities, where orbit averaging becomes less accurate, due to the large difference between orbital velocities at periastron and apastron. Possibilities for evolution without orbit averaging include direct integration of the PN equations of motion, without using the quasi-Keplerian formulation, as in~\cite{Csizmadia:2012wy,Ireland:2019tao}; osculating orbits, as in~\cite{Loutrel:2018ydu,Fumagalli:2025rhc} (though these do not include spin contributions); or by extending the closed-form solution to the $1.5$PN non-dissipative dynamics from~\cite{Samanta:2022yfe} \edit{(recently extended to $2$PN in \cite{Colin:2026iry} modulo some orbital-timescale evolutions of the spins that are still $1.5$PN)} to incorporate the effect of radiation reaction (and possibly higher-order terms) using perturbation theory. Additionally, one will need further improvements to the accuracy of our hybrid evolution code for it to be below the statistical errors anticipated for third\edit{-}generation ground-based detectors like Einstein Telescope and Cosmic Explorer as well as space-based detectors like LISA, with their expected exquisite sensitivity.

\edit{By the time Einstein Telescope, Cosmic Explorer, or LISA are providing data, it may be possible to infer the spin distributions at formation directly using population-synthesis models parameterized by metallicity, common-envelope efficiency, black hole natal spins, and other parameters, as well as branching ratios between different formation channels. Such population-synthesis-based inference is already beginning to be explored in simpler settings~\cite{Andrews:2020pjg,wong2023backward,Colloms:2025hib}, only considering the masses and a few other parameters from the GW inference. In particular, only~\cite{Colloms:2025hib} considers the spins, and they just consider the (inspiral) effective spin parameter rather than the full $6$-dimensional precessing spin configuration. (Reference~\cite{wong2023backward} only uses the masses in their inference, while the other two references also include the binary's redshift.) Therefore, with the expected extensions of these analyses, it may not be necessary to use the tilts at infinity to perform population-synthesis inference for third-generation detectors, and one will instead be able infer the tilts at formation directly for a given population model, if those quantities are of interest. However, at present, directly evolving the spins from formation to a frequency in the band of ground-based GW detectors (or vice versa) is computationally completely infeasible using current orbit-averaged evolution codes even for some predictions for binaries from dynamical formation channels, as discussed in Appendix~\ref{app:pop_syn_evol}, with estimated computation times of more than a year for some binaries (and the computation times for isolated formation channels are even longer): Being able to evolve with orbit-averaged evolution in such cases would require a new approach to the evolution of precessing binaries, such as the extension of the results in~\cite{Samanta:2022yfe,Colin:2026iry} discussed above.}

\edit{One can still use hybrid evolution to evolve all binaries to formation in a computationally feasible manner. However, if using hybrid evolution, it is best to use the tilts at infinity as a common reference point for the analysis. That is, one would compute the tilts at infinity predicted by the population-synthesis model under consideration and compare those predictions to the observed tilts at infinity. Specifically, the tilts at infinity include all the information of the binary's precessional degrees of freedom that one can constrain using precession-averaged evolution. (To see this, recall that the tilts at infinity are all the information about the precessional degrees of freedom needed to initialize forward precession-averaged evolution, as discussed in~\cite{Gerosa:2015tea}.) If, instead, one evolved backwards to the binary's formation using hybrid evolution, then one would only obtain the tilts at formation up to the variation over a precessional cycle (from conservative evolution) at that point in the binary's evolution, while if one uses hybrid evolution to evolve forward from formation to the band of ground-based detectors, then one has to sample over the precessional phase at the transition to orbit-averaged evolution, as discussed in~\cite{Reali:2020vkf}. In general, the tilts at infinity provide a natural asymptotic reference point, and therefore a common space in which detector-band posterior samples and population-synthesis predictions can be compared. We illustrate this point in Appendix~\ref{app:tilts_at_infty_specific_formations}, where we compute the distributions of \(\cos\theta_{1,\infty}\) for selected isolated and dynamical formation scenarios, showing the expected differences.}

\edit{As we have illustrated in this paper,} the tilts at infinity are not always a good approximation to the tilts obtained when evolving back to $e \nearrow 1$ in eccentric cases, and that the tilts obtained when evolving backwards to a finite separation may be a better approximation, at least when evolving with orbit averaging. If this conclusion holds when evolving without orbit averaging, future work will generalize the bounds on the difference between the tilts at a high eccentricity and $e \nearrow 1$ that we derived here for the orbit-averaged case to the case without orbit averaging and (if possible) sharpen it. We will then use these bounds to determine where in parameter space it is better to compute the tilts at infinity as a proxy for the tilts at formation for eccentric, precessing BBHs and where one should just evolve backwards to a finite separation (and what separation to evolve backwards to, to obtain a given accuracy).

\acknowledgments

We thank Matthew Mould for useful comments on the manuscript. We also thank the authors of~\cite{Belczynski:2017gds} for allowing us to include the data from that paper we use in our data release~\cite{singh_et_al_2025_zenodo}. This data was originally released with that paper but is no longer publicly accessible. AS is supported by NSF grant PHY-2308887. NKJ-M is supported by NSF grant AST-2205920. AG is supported in part by NSF grants PHY-2308887 and AST-2205920. KSP acknowledges support from STFC grant ST/V005677/1. The authors are grateful for computational resources provided by the LIGO Laboratory and supported by National Science Foundation Grants PHY-0757058 and PHY-0823459.
This study used the software packages LALSuite~\cite{LALSuite}, matplotlib~\cite{Hunter:2007ouj}, numpy~\cite{Harris:2020xlr}, pandas~\cite{mckinney-proc-scipy-2010},
and scipy~\cite{Virtanen:2019joe}.

This is LIGO document number P2500251.

\appendix

\section{Evolution of binaries from specific formation scenarios}
\label{app:pop_syn_evol}

In this section, we show the differences in the cosines of spin tilts at formation and at $10$~Hz for two different types of binary formation, namely, dynamical formation (with several specific formation scenarios) and isolated formation. \edit{Our goal in this section is not to perform a statistically complete or a full population study, but rather to provide illustrative examples using publicly available datasets to demonstrate that significant differences between tilts at reference frequency and at formation can arise in both dynamical and isolated channels.} We choose $10$~Hz (rather than the $20$~Hz used elsewhere in this paper) because $10~\mathrm{Hz}$ is a standard lower bound for the design sensitivity of LIGO and Virgo~\cite{LIGOScientific:2014pky,VIRGO:2014yos}, coming from fundamental limits due to seismic and Newtonian noise. If we instead used $20$~Hz, we would have larger differences compared to the tilts at formation, so the differences we obtain for the spin tilts between formation and at $10$~Hz are conservative. We use the data from~\cite{Kritos:2022ggc} (as released in \cite{konstantinos_2024_10632717}) for the dynamical formation scenario (taking the randomly sampled spin angles to be at formation) and the data from~\cite{Belczynski:2017gds} (released in~\cite{singh_et_al_2025_zenodo}) for the isolated formation scenario.
Some properties of the sampled BBHs from both formation scenarios are summarized in Table~\ref{tab:bbh_form_chan}.

\begin{table*}

\centering
\caption{Properties of $660$ ($500$) sampled BBHs from dynamical (isolated) formation scenarios. Both of these scenarios lead to some binaries with significant spin misalignments at formation: The minimum value of $\cos\theta_A$ ($A\in\{1,2\}$) is $-0.998$ for the dynamical binaries, while it is $-0.876$ for the isolated binaries. \edit{For all these binaries, the merger happens within the age of the universe.}}
\label{tab:bbh_form_chan}
\begin{tabular}{c|c c|c c}
\hline
\hline
\multirow{2}{*}{\textbf{Parameter}} 
& \multicolumn{2}{c|}{\textbf{Dynamical Formation}~\cite{Kritos:2022ggc}} 
& \multicolumn{2}{c}{\textbf{Isolated Formation}~\cite{Belczynski:2017gds}} \\
\cline{2-5}
 & Range & Median & Range & Median \\
\hline
$M$ ($M_{\odot}$)
    & $[49,\,178]$ & $95$   
    & $[13,34]$ & $20$ \\

$q$
    & $[0.34,\,0.999]$ & $0.81$   
    & $[0.36,\,0.9995]$ & $0.78$ \\

$e_{\rm formation}$
    & $1-[2.0\times10^{-6},\,9.8\times10^{-1}]$
    & $1-(9.5\times10^{-3})$
    & $[0.004,\,0.80]$ & $0.14$ \\

$\chi_1$
    & $[0.287,\,0.747]$ & $0.686$
    & $[0.102,\,0.267]$ & $0.166$ \\

$\chi_2$
    & $[0.666,\,0.686]$ & $0.686$
    & $[0.094,\,0.142]$ & $0.101$ \\

\hline
\hline
\end{tabular}

\end{table*}

In the dynamical formation case, there are $1144$ binaries in the combined data (corresponding to all values of initial stellar number $N$ and virial radius $r_{v}$) from~\cite{Kritos:2022ggc} for which both spin magnitudes are nonzero. These are all binaries that are formed hierarchically from second generation (2G) or third generation (3G) black holes, since~\cite{Kritos:2022ggc} assumes that all first generation black holes have zero spin. We are able to successfully evolve a total of $660$ BBHs to $10$ Hz using the orbit-averaged evolution code~\cite{Phukon:2025yva} ($624$ are 2G-2G BBHs and $36$ are 2G-3G BBHs). The population properties at formation for these $660$ binaries are given in Table~\ref{tab:bbh_form_chan}. The other binaries would either have been computationally much more expensive to evolve to 10~Hz or failed to start their evolution because one of the internal consistency checks of the PN expressions failed for the initial conditions. Specifically, $481$ binaries with median initial eccentricity $0.93$ did not finish evolving in more than $14$ days, while the other binaries, with a minimum eccentricity of $0.97$ and median of $0.99$, evolved in at most $3$ days.\footnote{Recall that the coalescence time scales with initial eccentricity $e_0$ as $(1 - e_0^2)^{7/2}$ in the high-eccentricity limit~\cite{PhysRev.136.B1224}, so a decrease of $e_0$ from $0.99$ to $0.93$ corresponds (\emph{ceteris paribus}) to an increase of the coalescence time by a factor of $\sim 800$.} Along with this, $3$ binaries (with eccentricities higher than $0.99998$) failed to evolve because the internal PN consistency checks failed at the start of the evolution. Specifically, these binaries are unbound according to the PN energy expression used in the evolution. Among the binaries that we were able to evolve, $589$ of them are three-body induced binaries, $26$ are the remnant binaries from the ZLK process, and $45$ are from binary-single interactions.

For the isolated formation case, we randomly select $500$ binaries from the data from~\cite{Belczynski:2017gds}, where we use the M30 model (see Table~II in their paper). To ensure that we use the convention $m_1 > m_2$ consistently throughout the paper, we swap the parameters for the binaries  where $m_1 < m_2$ in the isolated binaries dataset. These binaries have significantly smaller eccentricities at formation than the dynamically formed binaries (see Table~\ref{tab:bbh_form_chan}), while the median semimajor axis of the isolated binaries is only about a fifth of that of the dynamically formed binaries. Thus, the isolated binaries take $\sim 10^{4}$ times longer to reach $10$~Hz than the dynamically formed binaries. This makes using only the orbit-averaged code completely impractical to evolve the isolated binaries to $10$~Hz, since in most cases we would expect the computation to take a substantial amount of time. Thus, for these isolated binaries, we use a slight modification of the quasicircular hybrid evolution code from~\cite{Johnson-McDaniel:2021rvv} to evolve forwards (including the sampling in the precessional phase at the transition from precession-averaged to orbit-averaged evolution as described in~\cite{Gerosa:2015tea}). Since the eccentricities at $10$~Hz obtained from the leading-order expression from Peters~\cite{PhysRev.136.B1224} are all less than $3.21 \times 10^{-6}$, we expect that the assumption of quasicircular evolution is quite accurate for these binaries. Nevertheless, the largest eccentricity at $v_\text{trans}$ is $3.74 \times 10^{-3}$. Thus, to check the accuracy of the quasicircular approximation, we did the following: For the binary with the largest eccentricity at $v_\text{trans}$, we terminated the precession-averaged evolution at the $L_{\rm trans}$ value given by $v_\text{trans}$ and computed the eccentricity at that point from the Peters expression. We then evolved from $v_\text{trans}$ to $10$~Hz using the eccentric orbit-averaged code. The difference in the cosines of the tilts at $10$~Hz for quasicircular and eccentric evolutions was $< 3 \times 10^{-7}$, thus showing that quasicircular evolution is an excellent approximation for these binaries.

\begin{figure}
\includegraphics[width=.90\linewidth]{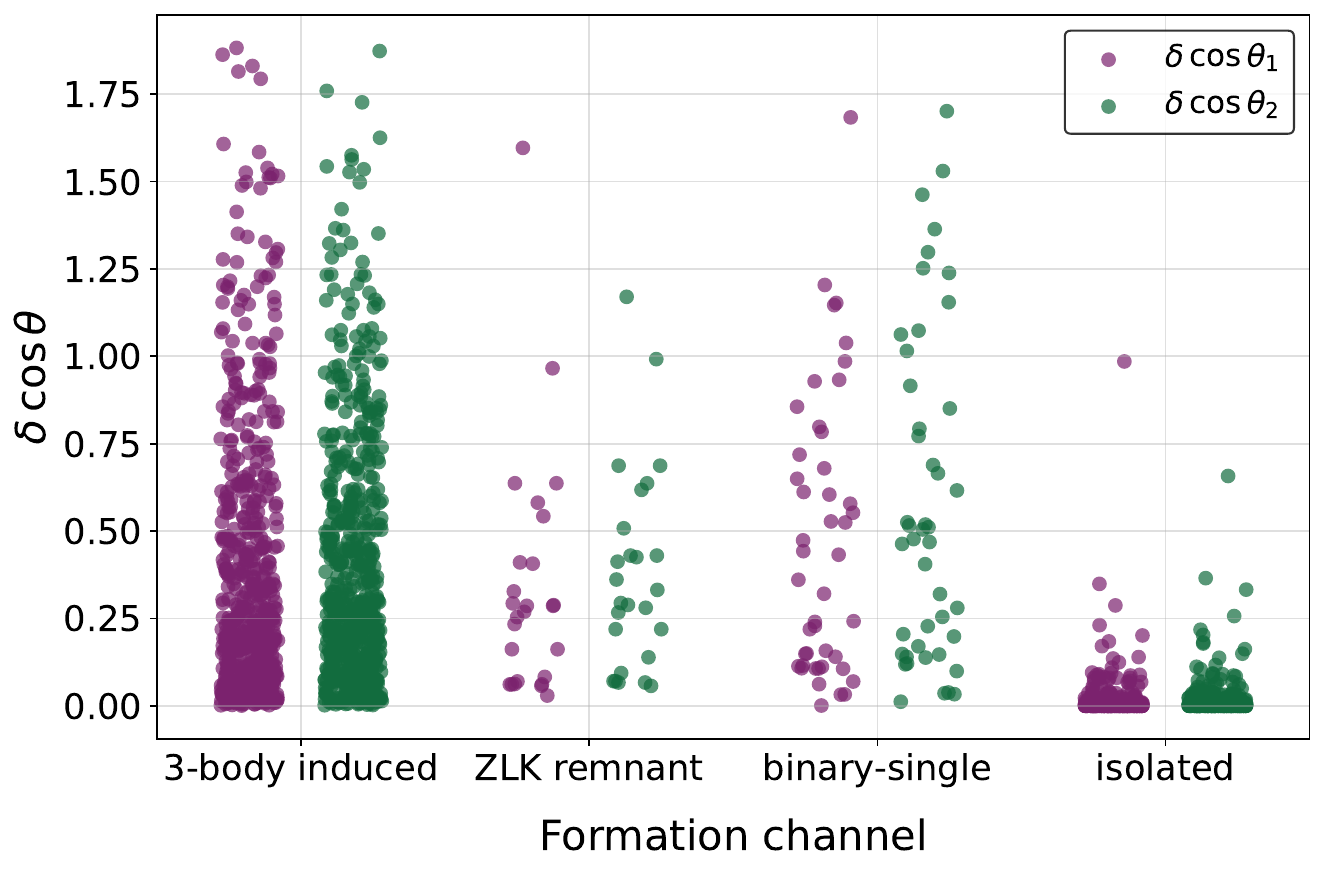}
\caption{Absolute values of the differences in cosines of spin tilts at formation and $10$~Hz for the $660$ (2G-2G and 2G-3G) BBHs from~\cite{Kritos:2022ggc} for three dynamical formation scenarios and $500$ randomly selection binaries from~\cite{Belczynski:2017gds} for the isolated formation scenario. We introduce a random horizontal jitter for each tilt's result to improve visualization. Out of $660$ dynamically formed binaries, a total of $509$ ($58$) binaries have differences greater than $0.1$ ($1$) in the cosines of tilt 1, while $534$ ($59$) binaries have differences greater than $0.1$ ($1$) in the cosines of tilt 2. Only $11$ isolated binaries have differences larger than $0.1$ in the cosines of tilt 1, while $14$ binaries have differences larger than $0.1$ for tilt 2. None of the isolated binaries have a difference larger than $1$ in either tilt.}

\label{fig:delta_cos_tilt_form_vs_ref_rapster_dynamical}
\end{figure}

We show the differences between the cosines of the tilts at formation and $10$~Hz in Fig.~\ref{fig:delta_cos_tilt_form_vs_ref_rapster_dynamical}. We observe that many of binaries, especially those formed through dynamical channels, show significant differences (greater than $0.1$) in the cosines of spin tilts between formation and at $10$~Hz, with even differences greater than unity in some cases. This shows that the tilts in the detector's sensitive band can be significantly different from those at the formation.

\section{\edit{Tilts at infinity for binaries from specific formation scenarios}}
\label{app:tilts_at_infty_specific_formations}

\begin{figure}
\includegraphics[width=.90\linewidth]{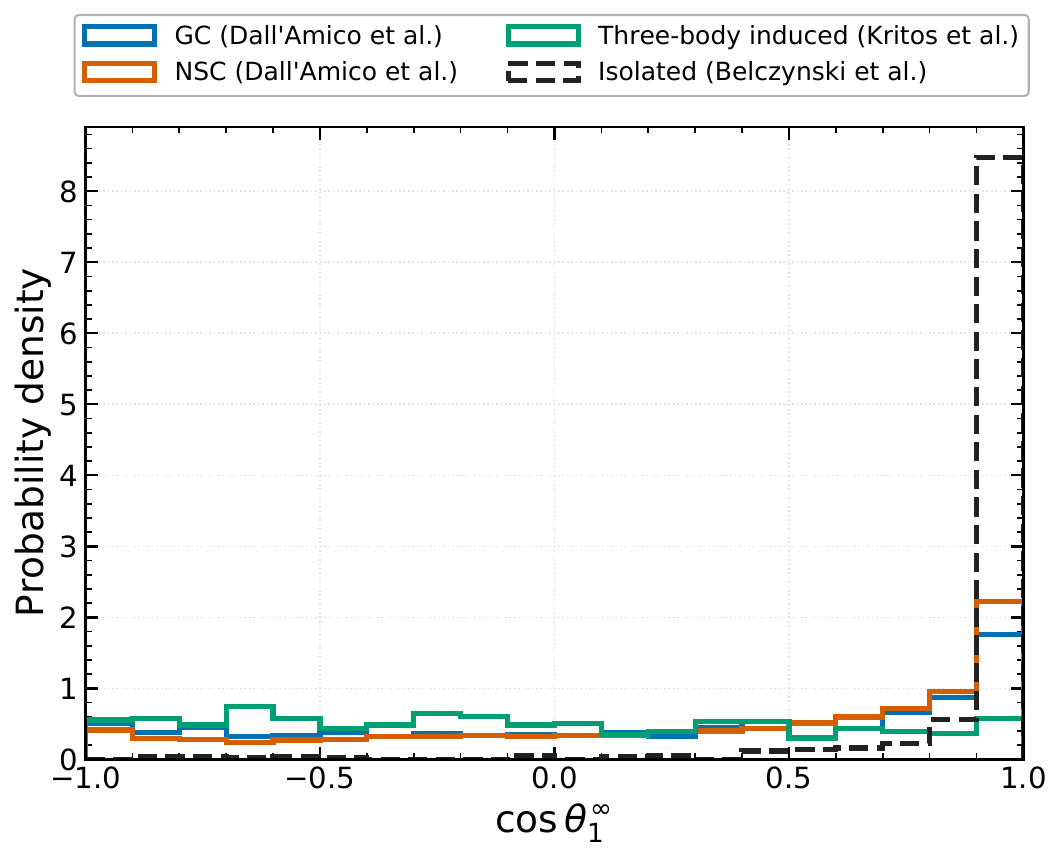}
\edit{\caption{Distributions of \(\cos\theta_{1}^{\infty}\) for selected dynamical and isolated formation scenarios. The dynamical samples include three-body-induced Rapster binaries from Kritos \emph{et al.}~\cite{konstantinos_2024_10632717} and GC and NSC binaries from Dall'Amico \emph{et al.}~\cite{DallAmico:2023neb}, while the isolated BBHs are taken from Belczynski \emph{et al.}~\cite{Belczynski:2017gds}. The distribution of \(\cos\theta_{2}^{\infty}\) is similar and thus not shown. The isolated binaries prefer spin tilts at infinity that are aligned with the orbital angular momentum, while the dynamical samples produce broader distributions. This illustrates that tilts at infinity can provide a fiducial reference point for distinguishing different population-synthesis models, avoiding the need to choose an arbitrary finite reference separation.}
\label{fig:cos_tilt1_at_infinity_selected_channels}
}
\end{figure}

\edit{In this section, we illustrate the difference in the distribution of tilts at infinity for different formation channels, thus showing how these distributions can help distinguish between different formation channels when using the tilts at infinity as a reference point, e.g., when it would only be computationally feasible to evolve the BBH spins from the band of ground-based GW detectors back to formation using hybrid evolution. We display the distributions of tilt~$1$ at infinity for BBHs from several formation scenarios in Fig.~\ref{fig:cos_tilt1_at_infinity_selected_channels}. (The distributions of tilt~$2$ at infinity are very similar, so we do not show them.) The first dynamical sample consists of the \(432\) three-body-induced binaries from the Rapster BBHs from Kritos \emph{et al.}~\cite{Kritos:2022ggc}, already used in Appendix~\ref{app:pop_syn_evol}. The other Rapster dynamical channels shown in Fig.~\ref{fig:delta_cos_tilt_form_vs_ref_rapster_dynamical}, namely the \(24\) remnant binaries from the ZLK process and the \(34\) binaries from binary-single captures, contain too few systems to allow good statistics and are therefore not included here.}

\edit{While the Rapster binaries just take the spins to be uniformly distributed on the $2$-sphere, and thus are also expected to have a uniform distribution of cosine tilts at infinity (up to statistical fluctuations), as we indeed find, there are other predictions from dynamical formation that predict a more complicated distribution of tilts at formation. Specifically, we consider the flyby cases from Dall'Amico \emph{et al.}~\cite{DallAmico:2023neb} (using the data released in~\cite{dall_amico_2023_7684085}). Since the data release only provides the coalescence time from the beginning of the simulation and not the binary's separation or frequency at the end of the simulation, which we need to compute the tilts at infinity, we use the coalescence time to obtain the binary's semimajor axis at an approximation to the end of the simulation using the Peters expression~\cite{PhysRev.136.B1224}. To make sure that this is a good approximation, we thus require that the coalescence time is larger than \(10^8\,{\rm yr}\), so that the duration of the simulation in~\cite{DallAmico:2023neb} (at most $10^6$~yr) is negligible compared to the subsequent inspiral time. This gives \(1708\) binaries formed in globular clusters (GC) and \(8319\) binaries formed in nuclear star clusters (NSC). It also leaves \(72\) young-star-cluster binaries, which we do not plot, since this is too few to provide good statistics. We use the orbital-plane tilts at the end of the simulation to compute the binary's initial spin tilts, assuming that both black hole spins are aligned with the binary's orbital angular momentum before the flyby. Thus, the tilts after the flyby are equal to the orbital-plane tilt and $\phi_{12} = 0$. Since the spin magnitudes are not provided in the data release, and are not used in the computation of the orbital-plane tilt, we draw them from the Maxwellian spin-magnitude distribution used in~\cite{DallAmico:2023neb}.}

\edit{Finally, we include the same set of \(500\) isolated binaries from Belczynski \emph{et al.}~\cite{Belczynski:2017gds} used in Appendix~\ref{app:pop_syn_evol}. We compute the tilts at infinity starting from tilts at $10$ Hz using our hybrid evolution code for the dynamically formed binaries, to replicate what one would do in practice. For the isolated binaries, we use only the precession-averaged evolution from formation to infinity because full orbit-averaged evolution over this range is computationally prohibitive.} 

\edit{The resulting distributions of tilts at infinity for these formation channels illustrate that when one uses the spin tilts at infinity as a reference point, they can help distinguish between different formation channels, especially when one can only use hybrid evolution (because of computational constraints) to evolve the BBH spins from the band of ground-based GW detectors back to formation.}

\section{Example usage of code}
\label{app:example_usage}

Here, we illustrate the use of our eccentric hybrid evolution code with some examples. The function \texttt{calc\_tilts\_at\_infty\_hybrid\_evolve()} computes the tilts at infinity for a BBH using hybrid evolution. The mandatory inputs to this function are as follows (with the binary parameters all in SI units): $m^{}_1$, $m^{}_2$, $\chi^{}_{1}$, $\chi^{}_{2}$, $\theta^{}_{1}$, $\theta^{}_{2}$, $\phi^{}_{12}$, and $f_{\rm ref}$. Here, $m^{}_1$ and $m^{}_2$ represent the masses of primary and secondary black holes respectively, while $\chi_{1,2}$ represent the dimensionless spin magnitudes. The spin angles are represented by $\theta^{}_{1,2}$ (spin tilt angles) and $\phi^{}_{12}$ (the relative azimuthal angle), while $f_{\rm ref}$ denotes the GW frequency to start the evolution with, where this is defined in the orbit-averaged way discussed in~\cite{Phukon:2025yva}. Optional keyword arguments for the function include $\psi_{1}$, $e_{0}$ (the binary's eccentricity at $f_{\rm ref}$), \texttt{spinO}, and \texttt{phaseO}, where the default values for the first two are $0$ and for the second two are \texttt{None}, which corresponds to the maximum value available. Here \texttt{spinO} and \texttt{phaseO} denote twice the PN order of spinning terms and nonspinning terms, respectively, in the orbit-averaged evolution equations. For eccentric evolution, the only allowed value of \texttt{spinO} currently is $4$ (since one does not want the orbit-averaged evolution to have a lower PN order than the precession-averaged evolution), while \texttt{phaseO}${}\in \{0, 2, 3, 4, 5, 6\}$, so the default setting is $6$. There is an option to get detailed information on the evolution by passing \texttt{verbose=True}. In Table~\ref{tab:examples}, we give some examples of running this code to compute the spin tilts at infinity.

\begin{widetext}

\begin{center}
  \captionof{table}{Examples of using the code~\cite{amiteshgit} introduced in this paper to evolve eccentric, precessing BBHs, starting from a reference frequency of $20$ Hz, all the way to infinity, in an interactive Python session, using the highest PN order expressions available.}
  \label{tab:examples}
\end{center}

\vspace{-1em}

\lstset{language=Python}

\begin{lstlisting}
# Setup
>>> from lalsimulation.tilts_at_infinity import prec_avg_tilt_comp, calc_tilts_at_infty_hybrid_evolve
>>> from lal import MSUN_SI
>>> m1, m2 = 50., 45. # solar masses
>>> chi1, chi2 = 0.8, 0.6
>>> tilt1, tilt2, psi1, psi2 = 1.3, 0.4, 0.6, 2.7 # rad
>>> phi12 = psi2 - psi1 #relative azimuthal angle
>>> f0 = 20. # Hz
>>> e0 = 0.3        # starting eccentricity at 20 Hz

\end{lstlisting}

\vspace{-1em}

\begin{lstlisting}
# Calculate spin tilts at infinity
# For only precession-averaged evolution
>>>	prec_avg_tilt_comp(m1*MSUN_SI, m2*MSUN_SI, chi1, chi2, tilt1, tilt2, phi12, f0, e0)
{'tilt1_inf': 1.1104672438023913, 'tilt2_inf': 0.8510182154945696}

# The same through the hybrid evolution interface
>>> calc_tilts_at_infty_hybrid_evolve(m1*MSUN_SI, m2*MSUN_SI, chi1, chi2, tilt1, tilt2, phi12, f0, psi1, e0, prec_only=True)
{'tilt1_inf': 1.1104672438023913, 'tilt2_inf': 0.8510182154945696, 'tilt1_f_stop': None,
 'tilt2_f_stop': None, 'phi12_f_stop': None, 'f_stop': None, 'eccentricity_f_stop': None}

\end{lstlisting}

\vspace{-1em}

\begin{lstlisting}
# For hybrid evolution

>>> calc_tilts_at_infty_hybrid_evolve(m1*MSUN_SI, m2*MSUN_SI, chi1, chi2, tilt1, tilt2, phi12, f0, psi1, eccentricity=e0)
{'tilt1_inf': 1.1765971973318956, 'tilt2_inf': 0.7252032198568954, 'tilt1_f_stop': 1.276168597718065,
 'tilt2_f_stop': 0.47964766389584523, 'phi12_f_stop': 4.1728009459579205, 'f_stop': 0.0050440712586374776,
 'eccentricity_f_stop': 0.9960631317641313}

# Sometimes the orbit-averaged code will not be able to evolve to the desired empirical f_trans
# In this case, the precession-averaged evolution starts from the frequency at which the orbit-averaged code stopped.

>>> e0 = 0.001  # Eccentricity for this test case

# Use verbose=True to get more details of the evolution
>>> calc_tilts_at_infty_hybrid_evolve(m1*MSUN_SI, m2*MSUN_SI, chi1, chi2, tilt1, tilt2, phi12, f0, psi1, eccentricity=e0, verbose=True)

RuntimeWarning: The evolution stopped at f_stop = 19.695466447044392 Hz. This is outside the allowed 1% tolerance around the empirical transition frequency 0.005044066816799267 Hz. Starting the precession-averaged evolution from f_stop. Accuracy in spin tilts at infinity will be reduced if the stopping frequency is significantly higher than the desired transition frequency.
The tilts at f_stop are: tilt1 = 1.30404788878423, tilt2 = 0.38488248446149187, phi12 = 2.106964977851078 and eccentricity is 1.4523807720004001e-05
{'tilt1_inf': 1.118011978527311, 'tilt2_inf': 0.8375968780052747, 'tilt1_f_stop': 1.30404788878423,
 'tilt2_f_stop': 0.38488248446149187, 'phi12_f_stop': 2.106964977851078, 'f_stop': 19.695466447044392,
 'eccentricity_f_stop': 1.4523807720004001e-05}

\end{lstlisting}

\end{widetext}

\section{Derivation of bound on change in tilts for large eccentricities}
\label{app:delta_theta_bound}

Here we derive a bound on the change in the cosines of the tilts when evolving with orbit-averaged evolution from a large eccentricity to $e \nearrow 1$. Specifically, using the precession equations for the binary's spin vectors $\mathbf{S}_A$ and orbital angular momentum unit vector $\mathbf{\hat{L}}$ used in the orbit-averaged code [Eqs.~(7) in~\cite{Phukon:2025yva}], as well as $\cos\theta_1 = \mathbf{\hat{S}}_1\cdot\mathbf{\hat{L}}$ and $M\omega = v^3$ (where circumflexes denote unit vectors), one finds that
\<
\begin{split}
\frac{d\cos\theta_1}{dt} &= \frac{3v^6}{2(1 - e^2)^{3/2}}\left(1 + \frac{m_1}{m_2} \,+\, \frac{Mm_1}{L}\chi_\text{eff}\right)\frac{m_2^2\chi_2}{M^3}\\
&\quad\times[\mathbf{\hat{S}}_1,\mathbf{\hat{S}}_2,\mathbf{\hat{L}}]
\end{split}
\?
and the same expression for tilt~$2$ with the swap $1 \leftrightarrow 2$ in the indices. Here $\chi_\text{eff} = (\mathbf{S}_1/m_1 + \mathbf{S}_2/m_2)\cdot\mathbf{\hat{L}}/M$ is the effective spin and $[\mathbf{\hat{S}}_1,\mathbf{\hat{S}}_2,\mathbf{\hat{L}}]=\mathbf{\hat{S}}_1\times\mathbf{\hat{S}}_2\cdot\mathbf{\hat{L}}$ denotes the scalar triple product. We now use the leading expression
\<
\frac{de^2}{dt} = -2\frac{m_1m_2}{M^3}\frac{v^8}{15(1 - e^2)^{5/2}}\left(304 + 121e^2\right)e^2
\?
to convert the time derivative to an $e^2$ derivative, giving
\<
\begin{split}
\frac{d\cos\theta_1}{de^2} &= -\frac{45(1 - e^2)}{4v^2\left(304 + 121e^2\right)e^2}\left(1 + \frac{m_2}{m_1} \,+\, \frac{Mm_2}{L}\chi_\text{eff}\right)\\
&\quad\times\chi_2[\mathbf{\hat{S}}_1,\mathbf{\hat{S}}_2,\mathbf{\hat{L}}].
\end{split}
\?
We also have, from the Peters relation between the semimajor axis $a$ and eccentricity~\cite{PhysRev.136.B1224},
\<
\frac{1}{v^2} = \frac{a}{M} = \zeta\frac{e^{12/19}}{1 - e^2}\left(1 + \frac{121}{304}e^2\right)^{870/2299},
\?
where $\zeta$ is set by the value of $v$ at some point in the evolution. Thus, we have
\<
\begin{split}
\frac{d\cos\theta_1}{de^2} &= -\frac{45 \zeta}{1216e^{26/19}}\left(1 + \frac{121}{304}e^2\right)^{-1429/2299}\\
&\quad\times\left(1 + \frac{m_2}{m_1} \,+\, \frac{Mm_2}{L}\chi_\text{eff}\right)\chi_2[\mathbf{\hat{S}}_1,\mathbf{\hat{S}}_2,\mathbf{\hat{L}}].
\end{split}
\?
We now specialize to $e^2\in[e^2_\text{trans},1]$ and also take the maximum values of $(m_1\chi_1 + m_2\chi_2)/M$ for $|\chi_\text{eff}|$ and $1$ for $|[\mathbf{\hat{S}}_1,\mathbf{\hat{S}}_2,\mathbf{\hat{L}}]|$ to simplify the upper bound, at the cost of some slight weakening. We also note that $L$ increases with $e$ from the Peters expression. We then obtain
\<
\begin{split}
\left|\frac{d\cos\theta_1}{de^2}\right| &\leq \frac{45 \zeta}{1216e_\text{trans}^{26/19}}\left(1 + \frac{121}{304}e_\text{trans}^2\right)^{-1429/2299}\\
&\quad\times\left[1 + \frac{m_2}{m_1} + \frac{m_2(m_1\chi_1 + m_2\chi_2)}{L_\text{trans}}\right]\chi_2.
\end{split}
\?
Moreover, we have
\<
\zeta = \frac{1 - e_\text{trans}^2}{v_\text{trans}^2e_\text{trans}^{12/19}}\left(1 + \frac{121}{304}e_\text{trans}^2\right)^{-870/2299},
\?
so an upper bound on the change in $\cos\theta_1$ between $e^2 = e^2_\text{trans}$ and $e^2 = 1$ is given by
\<
\begin{split}
|\Delta\cos\theta_1| &= \left|\int_{e^2_\text{trans}}^1
\frac{d\cos\theta_1}{de^2}de^2\right|\\
&\leq (1 - e_\text{trans}^2)
\max_{e^2\in[e^2_\text{trans},1]}
\left|\frac{d\cos\theta_1}{de^2}\right|\\
&\leq
\frac{45(1 - e_\text{trans}^2)^2}
{(1216 + 484e_\text{trans}^2)e_\text{trans}^2v_\text{trans}^2}\\
&\quad\times
\left[
1 + \frac{m_2}{m_1}
+ \frac{(m_1\chi_1 + m_2\chi_2)v_\text{trans}}
{m_1(1 - e_\text{trans}^2)^{1/2}}
\right]\chi_2.
\end{split}
\?
where we have used Eq.~\eqref{eq:L} to express $L_\text{trans}$ in terms of $v_\text{trans}$ and $e_\text{trans}$.
(The analogous expression for $|\Delta\cos\theta_2|$ is obtained by taking $1 \leftrightarrow 2$ in the indices.)
This is not the sharpest upper bound possible, but suffices for our purposes. Of course, we have used the leading-order Peters expression for the evolution of $a$ with $e$ as well as for $de^2/dt$, while the orbit-averaged evolution includes higher-PN corrections. However, for the bounds we compute in Sec.~\ref{sec:inf_vs_form}, these corrections are not large. In particular, noting that $\bar{x}$ decreases with $e$ from the Peters expression, the $1$PN corrections to $d\omega/dt$ and $de^2/dt$ that would give the $1$PN corrections to the Peters expression are fractionally less than a $3\%$ correction to the leading-order contributions (with the larger contribution for binary~B, with its higher eccentricity). Thus, they are not a large correction to the bound we derived. We leave deriving a strict bound on the corrections from higher-PN terms to future work.

\vspace{1cm}

\bibliography{ecc_hyb_evol_refs}

\end{document}